\pdfoutput=1
\PassOptionsToPackage{protrusion, expansion}{microtype} 

\documentclass[sigconf,nonacm, pdfa]{acmart}

\usepackage[a-2b]{pdfx}
\usepackage{adjustbox}
\usepackage{lineno}
\usepackage{algorithm}
\usepackage{algorithmic}
\usepackage{multirow}
\usepackage[tight,footnotesize]{subfigure}
\usepackage{epsfig}

\usepackage{amssymb}
\usepackage{footmisc}
\usepackage{footnpag}
\usepackage{sidecap}
\usepackage{amsmath}
\usepackage{paralist}
\usepackage{color}
\usepackage{perpage}
\usepackage{verbatim}
\usepackage{enumitem}
\usepackage{tikz}
\usepackage{listings}
\usepackage{graphicx}
\usepackage{soul}
\usepackage{fancyhdr}
\usepackage[inkscapeformat=png]{svg}
\usepackage{stfloats}
\usepackage{balance}
\usepackage{threeparttable}
\usepackage{multicol}

\newcommand\vldbavailabilityurl{https://github.com/JLiu-1/QPET-Artifact}
\newcommand\vldbpagestyle{empty}

\setlist{
    ,nosep
    ,itemindent=0pt
    ,leftmargin=*
    ,before=\vspace{.25\baselineskip}
    ,after=\vspace{.25\baselineskip}
    }

\begin{document}
\title{QPET: A Versatile and Portable Quantity-of-Interest-Preservation Framework for Error-Bounded Lossy Compression}


\author{Jinyang Liu}
\authornote{Both authors contributed equally to this research.}
\affiliation{%
  \institution{University of Houston}
  \city{Houston}
  \state{TX}
  \country{USA}}
\email{jliu217@central.uh.edu}

\author{Pu Jiao}
\authornotemark[1]
\affiliation{%
  \institution{University of Kentucky}
  \city{Lexington}
  \state{KY}
  \country{USA}}
\email{pujiao@uky.edu}

\author{Kai Zhao}
\affiliation{%
  \institution{Florida State University}
  \city{Tallahassee}
  \state{FL}
  \country{USA}}
\email{kzhao@cs.fsu.edu}

\author{Xin Liang}
\authornote{Corresponding author: Xin Liang, Department of Computer Science, University of Kentucky, Lexington, KY 40506.}
\affiliation{%
  \institution{University of Kentucky}
  \city{Lexington}
  \state{KY}
  \country{USA}}
\email{xliang@uky.edu}

\author{Sheng Di}
\affiliation{%
  \institution{Argonne National Laboratory}
  \city{Lemont}
  \state{IL}
  \country{USA}}
\email{sdi1@anl.gov}

\author{Franck Cappello}
\affiliation{%
  \institution{Argonne National Laboratory}
  \city{Lemont}
  \state{IL}
  \country{USA}}
\email{cappello@mcs.anl.gov}

\begin{abstract}
Error-bounded lossy compression has been widely adopted in many scientific domains because it can address the challenges in storing, transferring, and analyzing unprecedented amounts of scientific data. 
However, general error-bounded lossy compressors may fail to meet additional quality requirements for downstream analysis, a.k.a. Quantities of Interest (QoIs).
This may lead to uncertainties and even misinterpretations in scientific discoveries, significantly limiting the use of lossy compression in practice. 
In this paper, we propose QPET, a novel, versatile, and portable framework for QoI-preserving error-bounded lossy compression, which overcomes the challenges of modeling diverse QoIs by leveraging numerical strategies. 
QPET features (1) high portability to multiple existing lossy compressors, (2) versatile preservation to most differentiable univariate and multivariate QoIs, and (3) significant compression improvements in QoI-preservation tasks. 
Experiments with six real-world datasets demonstrate that integrating QPET into state-of-the-art error-bounded lossy compressors can gain 2x to 10x compression speedups of existing QoI-preserving error-bounded lossy compression solutions, 
up to 1000\% compression ratio improvements to general-purpose compressors, and up to 133\% compression ratio improvements to existing QoI-integrated scientific compressors.
\end{abstract}

\maketitle
\pagestyle{\vldbpagestyle}


\vspace{-3mm}
\ifdefempty{\vldbavailabilityurl}{}{
\vspace{.3cm}
\begingroup\small\noindent\raggedright\textbf{PVLDB Artifact Availability:}\\
The source code, data, and/or other artifacts have been made available at \url{\vldbavailabilityurl}.
\endgroup
}
\vspace{.3cm}

\section{Introduction}
\label{sec:intro}
Today's high-performance computing facilities and high-resolution instruments produce vast amounts of scientific data that overwhelm the data storage and transmission systems. 
According to recent studies, climate simulations generate hundreds of TB of data every 16 seconds~\cite{foster2017computing}, and fusion applications may produce 1 EB of data in a single run with current- and next-generation exascale computing systems~\cite{fusion}. In the molecular dynamics (MD) domain, the EXAALT project produces trajectory data with a trillion time steps~\cite{exaalt}, which requires hundreds of terabytes of disk space in a scientific data management system (e.g., HDF5~\cite{hdf5}).
Those examples pose grand challenges in the underlying data management tasks, including I/O and data transfer, which necessitate effective data compression. 

Data compression has been widely adopted in databases and data management systems for reducing data volumes~\cite{h264,h265,gorilla,jensen2018modelardb} and accelerating queries~\cite{chen2001query, arion2007xquec, zhang2018efficient, zhang2021g}. 
Nonetheless, challenges arise when applying existing compression techniques in the context of scientific data. 
On the one hand, while lossless compression techniques~\cite{gorilla,alakuijala2018brotli,lindstrom2017fpzip,zstd,nvcomp} can recover data to the exact precision, they suffer from limited compression ratios in floating-point scientific data. 
On the other hand, lossy compressors for natural media data and time-series databases, such as JPEGs \cite{jpeg,jpeg2000}, H.26x series~\cite{h264,h265,h266}, and SummaryStore~\cite{summarystore} cannot provide a quantifiable bound on the decompression errors, leading to uncertainties in the downstream data visualization and analytics. 
To this end, error-bounded lossy compression has been proposed as a viable way to reduce the data size while providing guaranteed error control. 
For instance, ModelarDB~\cite{jensen2018modelardb,kejser2019scalable} is a typical example of time-series databases that support error-bounded lossy compression.

The same trend is observed in scientific data management systems.
Over the past years, several error-bounded lossy compressors~\cite{Xin-bigdata18,sz3,zfp,SPERR,HPEZ,cusz,cusz+} have been developed and adopted by multiple scientific domains, as they can effectively accelerate data management in scientific applications.
These compressors feature strict error control based on user requirements, so domain scientists can control the impact of lossy compression according to specific needs. 
However, most compressors only provide error bounds on the raw data, even though the scientists care more about the errors in their derived quantities or downstream analyses.  
Such derived information, also known as Quantities of Interest (QoIs), is of utmost importance to the fidelity and integrity of scientific discoveries. 
Typical QoIs include symbolistic functions~\cite{jiao2022toward}, derivatives and gradients~\cite{su2022understanding}, integrals~\cite{mgardqoi2}, and topological features~\cite{liang2022toward, yan2023toposz,xia2024preserving}.
Failure to quantify the distortions in QoIs may lead to misinterpretation of the data and even falsified discoveries, significantly limiting the use of error-bounded lossy compressors in practice. 

A few QoI-preserving compressors have already been proposed and applied in scientific applications to bridge the gap. 
For instance, MGARD~\cite{mgardqoi} has enabled the support for bounded linear QoIs, which is further integrated into fusion applications to preserve critical derived quantities~\cite{gong2021maintaining}; variations of SZ have been proposed to tackle the preservation of symbolistic QoIs~\cite{jiao2022toward}, contour tree~\cite{yan2023toposz}, and critical points in vector fields~\cite{liang2022toward, xia2024preserving}. 
Despite those efforts, several significant challenges remain unaddressed in the context of QoI reservation. 
(1) Existing QoI-preserving lossy compressors are finely crafted for particular QoIs, which lose generalization and extensibility to more diverse QoI formats. (2) Most existing methods rely on custom point-wise error bounds to enable the preservation of QoIs, but effectively determining or storing these error bounds is non-trivial. (3) Existing approaches usually tightly couple the QoI-preserving mechanism and compression pipeline, which fail to leverage newly developed compression methods.  
To address those challenges, we propose \textbf{QPET}, a \textbf{novel, versatile, and portable framework for efficiently preserving symbolistic QoIs}. 
In particular, we notice that various symbolistic QoIs can be modeled in mathematical formats composed of differentiable functions. Moreover, the numerical approximations of the QoIs with a simplified format (e.g., Taylor expansion with second-order derivatives) can provide a unified, efficient, and well-performing algorithmic routine to determine the point-wise error bounds. 
Thus, we develop a highly optimized QoI-preserving framework that can easily adapt to diverse QoIs and scientific lossy compression pipelines. 
In summary, our contributions are three-fold:

\begin{itemize}
    \item  We propose QPET, a QoI-oriented Point-wise Error-bound auto-tuning framework to enable the preservation of diverse QoIs in scientific applications. In particular, QPET leverages numerical and probabilistic methods to derive sufficient error bounds on each data point, which easily adapts to most differentiable QoIs. 
    \item QPET effectively determines the best-fit point-wise error bound for each data value in the QoI-preserving compression. It also jointly optimizes the error-bound storage overhead and the reduction in raw data by a dynamic global error bound selection. 
    \item We integrate QPET into 3 state-of-the-art error-bounded lossy compressors. Experimental results demonstrate that QPET yields significant compression ratio (CR) and throughput gains in QoI-preserving compression tasks. Under the same QoI error threshold, QPET-integrated compressors achieve up to 1000\% CR improvements over the general-purpose compressors and up to 133\% CR improvements over existing QoI-preserving lossy compressors. It also achieves 2x to 10x compression speedups over existing QoI-preserving scientific lossy compression solutions.
\end{itemize}

We organize the rest of this paper as follows: Section~\ref{sec:related} discusses the related work. In Section~\ref{sec:background}, we formulate and clarify our research motivation and target. In Section~\ref{sec:framework}, we propose the high-level design of QPET. Section~\ref{sec:qoi} demonstrates the detailed designs and algorithms in QPET. In Section~\ref{sec:evaluation}, we present the evaluations of QPET. Section~\ref{sec:conclusion} concludes this work and discusses future plans.

\section{Related Work}
\label{sec:related}

\textbf{Traditional compression techniques for database systems:}
Various data compressors have been proposed for large databases in diverse data domains/formats. 
Generic lossless compressors such as GZIP~\cite{gzip}, ZSTD~\cite{zstd}, and LZ4~\cite{lz4} are widely used to reduce storage requirements~\cite{h264,h265,gorilla,jensen2018modelardb} and accelerate queries~\cite{chen2001query, arion2007xquec, zhang2018efficient, zhang2021g}. 
Several lossless compressors have also been proposed to deal specifically with the increasing amount of floating-point data.
A typical example is ndZip~\cite{ndzip,ndzip-gpu}, which leverages Lorenzo prediction~\cite{ibarria2003out} and residual encoding to compress floating-point data with high efficiency. 
To handle different types of data, Gorilla~\cite{gorilla} and Chimp~\cite{liakos2022chimp} have been designed for time-series data, and Buff~\cite{buff} has been tailored to process low-precision data.  
Despite their ability to recover the exact data, lossless compressors exhibit low compression ratios, severely limiting their use in practice.

To address the limited compression ratios in lossless compressors, lossy compression has been proposed as an alternative way to reduce data in database systems. 
Regarding lossy database compression methods, ModelarDB \cite{jensen2018modelardb,kejser2019scalable} and summaryStore \cite{summarystore} are designed for time-series data, which leverage data-processing techniques including PMC-mean~\cite{pmc-mean} and the linear Swing model\cite{swing}. 

\textbf{Emerging lossy compression techniques for scientific data:} To process the high volumes of scientific data while preserving the data quality for accurate post hoc analytics, scientific data compressors require not only decent compression ratios but also strict error control, where error-bounded lossy compression is becoming a promising option. 
It allows users to specify an error bound as a parameter and ensures the element-wise difference between the original data and decompressed data is less than the error bound. 

SZ3~\cite{szinterp,sz3} is one of the most widely used error-bounded lossy compressors with a prediction-based design. 
It utilizes dynamic spline interpolations to approximate the data, followed by linear-scaling quantization ~\cite{sz17} and lossless encoding to reduce data size with guaranteed error control. 
Later, QoZ/HPEZ~\cite{qoz,HPEZ} is proposed to significantly improve the quality of SZ3 via more complex interpolation schemes and quality-oriented auto-tuning mechanisms. 
Another category of error-bounded lossy compressors relies on domain transform for decorrelation. 
For instance, ZFP~\cite{zfp} is a fast compressor using near-orthogonal block transform, and SPERR~\cite{SPERR} utilizes more complex wavelet transforms to obtain better compression quality at the cost of lower compression throughput.
While error-bounded lossy compressors have guaranteed data error control, they cannot provide a quantifiable error bound on the downstream QoIs, as the QoIs could be highly diverse and non-linear.

\textbf{QoI-preserving lossy compression methods:} Several solutions have been proposed to enable error control for various QoIs. 
MGARD~\cite{mgardqoi} is one of the first compressors to enable QoI preservation, but it can only provide guaranteed error control for bounded-linear QoIs. 
In~\cite{liang2020toward}, cpSZ has been proposed to preserve the locations and types of critical points in piece-wise linear vector fields. 
In particular, it carefully derived sufficient point-wise error bounds based on the target QoI and leveraged them to guide the compression procedure for effective QoI preservation. 
This idea is further extended to cover the preservation of critical points in multilinear vector fields~\cite{liang2022toward}, critical points extracted by sign-of-determinant predicates~\cite{xia2024preserving}, contour trees~\cite{yan2023toposz}, and symbolistic QoIs~\cite{jiao2022toward}.

Most existing QoI-preserving lossy compressors suffer from low generalizability and adaptability. In particular, they feature specific error control mechanisms toward the target QoI, which is hard to generalize to new sets of QoIs. In addition, most existing works are tightly coupled to a specific compression pipeline, making it difficult to integrate with newly developed compression methods. 
In this work, we propose a portable module that can easily adapt to diverse QoIs and compression methods. This will significantly improve the efficiency and usability of QoI-preserving lossy compression. 
\section{Background}
\label{sec:background}
In this section, we introduce the background for the proposed work. In particular, we define the target QoIs to preserve, followed by a formulation of the QoI-preserving lossy compression problem. 

\subsection{Quantities of Interest in Scientific Data}
\label{sec:targetqoi}
As previously discussed, scientific applications usually require the preservation of QoIs to ensure the integrity of scientific discoveries if lossy compression is used. 
QoIs represent any derived information that is computed from the raw data, including physical properties~\cite{gong2021maintaining,wu2024error}, derivatives~\cite{su2022understanding}, and topological features~\cite{yan2023toposz}. 

In this work, we target the preservation of symbolistic QoIs 
, which can be formulated as mathematical functions of the input data.
This covers various QoIs in scientific domains, including kinetic energy in cosmology, density in fusion energy science, and vector magnitude in climatology. 
We list the categorizations of target QoIs that the proposed framework can preserve as follows:

\textbf{Univariate QoIs:} Any second-order differentiable QoI functions, particularly the elementary functions, which are the combination of the following basic functions:
\begin{itemize}
    \item Polynomial functions such as $x^2$, $x^3$, and $x^2+x+1$;
    \item Exponential functions such as $2^x$ and $e^x$;
    \item Logarithm functions such as $\log_2{x}$ and $\ln{x}$;
    \item Generalize power functions such as $(x+c)^{-1}$(i.e., $\frac{1}{x+c}$);
\end{itemize}
Composite elementary functions such as Sigmoid $\sigma(x) = \frac{1}{1+e^{-x}}$ and Hyperbolic Tangent $\tanh{x} = \frac{e^x-e^{-x}}{e^x+e^{-x}}$ can also be supported. 

    \textbf{Multivariate QoIs:} Any differentiable multivariate functions in the format of $F(x_1, x_2, ...x_n)$, in which $\{x_1, x_2, ...x_n\}$ are a set of data points, are also supported by QPET. The point set can be a data region or a vector data entry. Typical examples include:

\begin{itemize}
    \item Regional weighted sum of univariate QoI ($\frac{1}{n}\sum x^2$);
    \item Velocity scalar of 3-D velocity vector ($\sqrt{v_x^2+v_y^2+v_z^2}$);
    \item Kinetic energy of 3-D velocity vector ($\frac{1}{2}m\left(v_x^2+v_y^2+v_z^2\right)$);
\end{itemize}

While several existing works managed to preserve certain subsets of the aforementioned QoIs~\cite{jiao2022toward,mgardqoi}, 
they cannot be generalized to other QoIs easily because they heavily rely on the analytical form of the QoIs to make analysis.

\subsection{QoI-preserving Lossy Compression}
\label{sec:problem}
We mathematically define QoI-preserving lossy compression as providing guaranteed error control in both the downstream QoIs and the raw data, which is similar to existing works~\cite{jiao2022toward,mgardqoi,liang2022toward}.

Given an input data $X = \{x_i\}$, and a pair of compressor and decompressor $(\operatorname{Cmp}, \operatorname{Dec})$ that process $X$ to compressed data $C = \operatorname{Cmp}(X)$ and decompressed data $D = \{d_i\} = \operatorname{Dec}(C)$, for a QoI function $Q$ (which maps $X$ to another set of values $Q(X)$), a data error bound $\epsilon$, and a QoI error threshold $\tau$, in QoI-preserving error-bounded lossy compression, we require the errors of data and QoI are both bounded, i.e., $\left\|X-D \right\|_{\infty} \leq \epsilon$ and $\left\|Q(X)-Q(D) \right\|_{\infty} \leq \tau$.

Our research would like to optimize $\operatorname{Cmp}$ and $\operatorname{Dec}$ to maximize the compression ratio.
Because most QoIs are non-linear, setting a uniform error bound on each data point will result in divergent QoI errors. 
To address this issue, existing approaches~\cite{liang2020toward, liang2022toward, jiao2022toward} apply different point-wise error bounds on each data point to achieve effective QoI preservation. 
Nonetheless, these methods focused more on deriving the error bounds and overlooked their storage overhead, leading to suboptimal overall compression ratios. 
To optimize this strategy, selecting the best-fit point-wise error bound (composing a set of $\{\epsilon_i\}$) for each data point such that $|x_i-x_i^{\prime}| \leq \epsilon_i$ to maximize compression ratio, we need to jointly optimize the compressed data size ($|\operatorname{Cmp}_{\{\epsilon_i\}}(X)|$) and the compressed error-bound size of $\{\epsilon_i\}$ themselves ($\left|\operatorname{Cmp}(\{\epsilon_i\})\right|$). This optimization problem is shown in Eq.~\ref{eq:ebopt} and will be specified in Section~\ref{sec:qoi}.

\begin{equation}
\label{eq:ebopt}
\begin{split}
&\{\epsilon_i\} = arg\,max_{\{\epsilon_i\}}{\frac{|X|}{|\operatorname{Cmp}_{\{\epsilon_i\}}(X)|+|\operatorname{Cmp}(\{\epsilon_i\})|}} \\
s.t. \ &\|X-D\|_{\infty} \leq \epsilon \ and \ \left\|Q(X)-Q(D) \right\|_{\infty} \leq \tau \\
\end{split}
\end{equation}

\section{QPET Design Overview}
\label{sec:framework}
In this section, we provide an overview of the proposed QPET (QoI-oriented Point-wise Error-bound Tuning) framework, describing its modules, underlying algorithm, and its integration into several existing error-bounded lossy compression pipelines. 

\subsection{QPET composition and algorithm}
\label{sec:composition}
\begin{figure}[ht]
  \centering
  \raisebox{-1mm}{\includegraphics[scale=0.5]{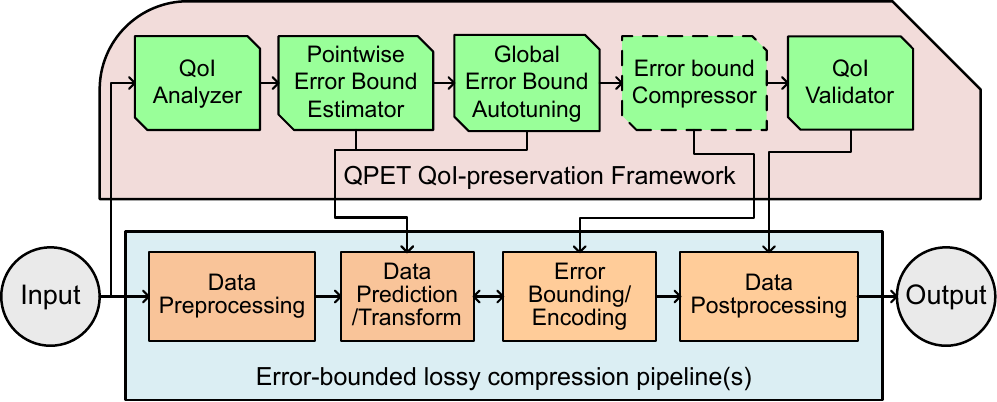}}
  \caption{QPET framework and integration}
  \label{fig:framework}
\end{figure}

In Figure \ref{fig:framework}, we demonstrate the composition of the QPET framework that provides QoI-preservation functionality, presenting how it integrates and interacts with general modular scientific error-bounded lossy compression pipelines. 
Compared to existing QoI-preserving work~\cite{jiao2022toward}, QPET introduces several brand-new or improved strategies to cover a wide range of QoI constraints and deliver high compression ratios. The components of QPET are:
\begin{itemize}
    \item \textbf{QoI Analyzer:} This symbolic module creates interfaces for efficient evaluations of the QoI function and its derivatives. 
    \item \textbf{Point-wise Error Bound Estimator:} To constrain the QoI error, this module computes an estimated best-fit compression error bound separately for each input data point.
    \item \textbf{Global Error Bound Auto-tuner:} This module auto-tunes the best-fit global data error bound for the compression. 
    \item \textbf{Error Bound Compressor:} This module compresses and stores the point-wise error bounds (will be skipped if the base compression pipeline cannot support point-wise error bounds) with improvements from \cite{liang2020toward,jiao2022toward}.
    \item \textbf{QoI Validator:} This module verifies the QoI constraints during online data prediction-quantization processes and corrects out-of-constraint data points by lossless storage. 
\end{itemize}

Leveraging our proposed QPET framework, Algorithm~\ref{alg:qoicomp} presents the process of QoI-preserving lossy compression. In Line 1, the QoI Analyzer analyzes input QoI and creates numerical interfaces for QoI/derivatives evaluations; In Line 5 and Line 10, the point-wise data error bounds are computed (detailed in Section~\ref{sec:univar} and ~\ref{sec:mulvar}); In Line 13, the global data error bound is auto-tuned (detailed in Section~\ref{sec:globalt}); In Line 17, the point-wise data error bounds are compressed and stored; In Line 18, the QoI validator module checks the decompressed QoI errors. The outlier data points that bring out-of-bound QoI errors will be losslessly encoded and appended to the compressed data (detailed in Section~\ref{sec:qoival}).

\begin{algorithm}[t]
\caption{\textsc{QoI error-constrained error-bounded lossy compression with QPET}} \label{alg:qoicomp} \footnotesize
\renewcommand{\algorithmiccomment}[1]{/*#1*/}
\begin{flushleft}
\textbf{Input}: input data $X = \left\{x_i\right\}$ of size $n$,  initial global error bound $\epsilon$, QoI function $Q$, QoI error threshold $t$ \\
\textbf{Output}: Compressed data $C$
\end{flushleft}
\begin{algorithmic} [1]
\STATE $\operatorname{QoIAnalysis}(Q)$ \COMMENT {Pre-analysis of QoI constraint} 
\STATE $\left\{\epsilon_i\right\} \gets \{\epsilon\}$ \COMMENT {Initialize point-wise error bounds} 
\IF [$F$ is univariate] {$\operatorname{isUnivariate}(Q)$} 
\FOR{$i = 1 \to n $}
\STATE $\epsilon_i \gets \operatorname{GetPointwiseEB\_UniVar}(x_i,Q,t,\epsilon)$ 
\COMMENT {Compute point-wise data error bounds by Algorithm~\ref{alg:point-wise}}
\ENDFOR
\ELSE [$Q$ is multivariate]
\FOR[$n_r$ is number of data regions ]{$i = 1 \to n_r $}
\STATE $R_i \gets \operatorname{GetDataRegion}(i)$
\STATE $ E_i \gets \operatorname{GetPointwiseEB\_MulVar}(R_i,Q,t,\epsilon)$ 
\COMMENT {Compute point-wise data error bounds $E_i$ for data region $R_i$ by Algorithm~\ref{alg:regional}}
\ENDFOR
\ENDIF
\STATE $\epsilon_g \gets \operatorname{TuneGlobalEB}\left(\left\{\epsilon_i\right\}\right)$ \COMMENT{Auto-tune global error bound by Algorithm~\ref{alg:globaltune}}
\FOR{$i = 1 \to n $}
\STATE $\epsilon_i \gets \min \left(\epsilon_i,\epsilon_g\right)$ 
\COMMENT {Update point-wise data errors bound with new global error bound}
\ENDFOR
\STATE $X^{\prime}, C \gets \operatorname{Compress}\left(X,\left\{\epsilon_i\right\},\epsilon_g\right)$ \COMMENT{Compress input data with tuned error bounds. $C$ and $X^{\prime}$ are the compressed data and decompressed data.}
\STATE $X_o \gets \operatorname{QoIValidation}(X, X^{\prime}, Q, t)$ \COMMENT{Validate QoI errors and collect out-of-constraint data}
\STATE $C \gets C + \operatorname{Lossless\_Compress}(X_o)$ \COMMENT{Losslessly compress $X_o$ and append the compressed data to $C$}
\RETURN $C$
\end{algorithmic}
\end{algorithm}

\subsection{Integrating QPET into error-bounded compression pipelines}

Because the design of QPET is only based on error-bound estimation/tuning and post-hoc data validation/correction, it does not rely on certain data compression techniques and implementation details. As such, QPET has high portability and adaptability to many existing error-bounded lossy compression pipelines.
In this work, we have integrated QPET into three state-of-the-art scientific error-bounded lossy compressors with two archetypes: SZ3 \cite{szinterp,sz3}, HPEZ \cite{HPEZ}, and SPERR \cite{SPERR}. SZ3 and HPEZ are interpolation-prediction-based error-bounded lossy compressors, and SPERR is a wavelet-transform-based error-bounded lossy compressor.
In the integration of QPET to each data compressor, the error-bound estimation and tuning modules (according to the QoI function and error threshold) are added before the data compression process, and the data validation/correction module serves as a data post-processing step in the compression pipeline.

\section{QPET Design Details}
\label{sec:qoi}

In this section, we present the design details of QPET introduced in Section~\ref{sec:framework}. First, we discuss the core of QPET: providing optimized point-wise error bounds for the QoI-preserving lossy compression. Then, we explain how to auto-tune the global data error bound according to the computed point-wise error bounds to optimize the compression ratio. Last, we specify how to correct the outlier data values to strictly respect the QoI error threshold.

\subsection{Computing Point-wise Error Bounds with Univariate QoI}
\label{sec:univar}
Regarding the point-wise error-bound computation, we first visit the univariate QoI cases and then extend it to multivariate QoI functions. We formulate the optimization of the error-bounded lossy compression with univariate QoI as follows. 
For each data point $x_i$ in the input data, given a global error bound $\epsilon_g$, a QoI function $f$, and a QoI error threshold $t$, we need to find the optimized point-wise error bound $\epsilon_i$ of $x_i$ from the following problem:

\begin{equation}
\label{eq:pwqoi}
\begin{split}
&\epsilon_i = \max \epsilon  \\
s.t. \ & \epsilon \leq \epsilon_g, \\
and \ &  \left|x_i^{\prime}-x_i\right|\leq \epsilon \implies \left|f\left(x_i^{\prime}\right)-f(x_i)\right| \leq t \\
\end{split}
\end{equation}

\begin{table}[t]
\centering
\caption{Mathematical notations for QoI-preserving}
\label{tab:notations}
\small
\begin{tabular}{|c|c|}
\hline
\textbf{Symbol} & \textbf{Description} \\ \hline
$X$ ($\left\{x_i\right\}$)         & Input Data(set)                  \\ 
\hline
$x_i$         & Input data point                 \\ 
\hline
$X^\prime$ ($\left\{x_i^\prime\right\}$)         & Decompressed Data(set)                  \\ 
\hline
$x_i^\prime$         & Decompressed data point               
            \\ \hline
$\epsilon$                           & General data error bound                  \\ \hline
$\epsilon_g$                           & Global data error bound                  \\ \hline
$\epsilon_i$ ($\left\{\epsilon_i\right\}$)                          & Pointwise data error bound (set)                 \\ \hline
$Q$                            & General QoI function                \\ \hline
$f$                            & Univariate QoI function                \\ \hline
$F$                            & Multivariate QoI function          \\ \hline
$\tau$         & General QoI error threshold                  \\ 
\hline
$t$         & Univariate QoI error threshold                \\ 
\hline
$T$         & Multivariate QoI error threshold                \\ 
\hline

\end{tabular}
\end{table}

Eq.~\ref{eq:pwqoi} guarantees that compressing $x_i$ with error bound $\epsilon_i$ can bound both the global data and QoI errors with required thresholds. Meanwhile, it is maximized to optimize the data compression ratio. We can optimize this compression task if we can solve $\epsilon_i$ effectively from Eq.~\ref{eq:pwqoi}. When the QoI function $f$ has a simple format, such as linear, quadratic, or simply logarithmic, Eq.~\ref{eq:pwqoi} may also have analytical solutions \cite{jiao2022toward}. However, analytical solutions have limitations: their existence and computational efficiency depend highly on the QoIs. Each analytical solution is customized for a separate QoI, and the computation method cannot be extended to generalized QoIs. To this end, our work proposes a numerical solution that provides a generalized and efficient approximation method for $\epsilon_i$ with a wide range of QoI formats. 
Specifically, as long as $f$ is second-order differentiable (any elementary function is infinitely differentiable), when computing $\epsilon_i$ for data point $x_i$, we will use its second-order Taylor expansion $f_2(x'_i) = f(x_i) + f^{\prime}(x_i)(x'_i-x_i)+\frac{f^{\prime\prime}(x_i)}{2}(x'_i-x_i)^2$ to estimate the QoI value at the decompressed data $x'_i$.
In particular, this Taylor expansion with the third-order remainder is:

\begin{equation}
\label{eq:sectaylor}
\begin{array}{l}
  f(x'_i) = f_2(x'_i) + \frac{f^{\prime\prime\prime}(\xi)}{3!}(x'_i-x_0)^3,
\end{array}
\end{equation}
where $\xi$ is a specific value in the interval $[x_i, x'_i]$ (or $[x'_i, x_i]$ if $x'_i < x_i$.
Given the error bound $\epsilon_i$ which ensures $x'_i \in [x_i-\epsilon_i, x_i+\epsilon_i]$ and assuming $f^{\prime\prime\prime}(x)$ is bounded in $[x_i-\epsilon_i, x_i+\epsilon_i]$ by $|f^{\prime\prime\prime}(x)| \leq M$,
the remainder $\frac{f^{\prime\prime\prime}(\xi)}{3!}(x-x_0)^3$ would be smaller than $\frac{M^3}{6}\epsilon_i^3$, i.e. $|f_2(x'_i)-f(x_i)| \leq \frac{M^3}{6}\epsilon_i^3$. 
With $f_2$, we can efficiently compute a good estimation for the best-fit $\epsilon_{i}$ by the following theorem:

\begin{theorem}
    \label{theo:1}
    Given a second-order differentiable QoI function $f$, a global error bound $\epsilon_g$, and a QoI error threshold $t$, for each data point $x_i$, if $f^{\prime\prime}(x_i) \neq 0$, an estimation for the best-fit data error bound $\epsilon_{x_i}$ to fit its QoI error threshold will be $\min{\left(\epsilon_g, \frac{\sqrt{|a|^2+2|b|t}-|a|}{|b|}\right)}$, in which $a = f^{\prime}(x_i)$ and $b = f^{\prime\prime}(x_i)$.
\end{theorem}

\begin{proof}
Using $f_2(x) = x_i + a(x-x_i)+\frac{b}{2}(x-x_i)^2$ as an approximation of $f$, we first find the maximum $\epsilon$ so that $|x_i^{\prime}-x_i| \leq \epsilon$ $\implies$ $|f(x_i^{\prime})-f(x_i)| \approx |f_2(x_i^{\prime})-f_2(x_i)| \leq |a(x_i^{\prime}-x_i)+\frac{b}{2}(x_i^{\prime}-x_i)^2| \leq t$. 
This is equivalent to maximizing $\epsilon$ which satisfies that $[-\epsilon, \epsilon] \subseteq A$, in which $A$ is the solution set of $|ax+\frac{b}{2}x^2| \leq t$. 

By solving $|ax+\frac{b}{2}x^2| \leq t$ (for simplicity we omit the detailed process), we can acquire that $\max{\epsilon} = \frac{\sqrt{|a|^2+2|b|t}-|a|}{|b|}$. Considering the global $\epsilon_g$, finally $\epsilon_{x_0} = \min{\left(\epsilon_g , \frac{\sqrt{|a|^2+2|b|t}-|a|}{|b|}\right)}$.
\end{proof}

Theorem ~\ref{theo:1} is valid when $b = f^{\prime\prime}(x_0) \neq 0$. Under $f^{\prime\prime}(x_0) = 0$, we have the following theorem as a supplement of Theorem ~\ref{theo:1}, whose proof is intuitively trivial:

\begin{theorem}
    \label{theo:2}
    Given a second-order differentiable QoI function $f$,  a global error bound $\epsilon_g$, and a QoI error threshold $t$, for a data point $x_0$ that $f^{\prime\prime}(x_0) = 0$, an estimation for the best-fit data error bound of $x_0$ will be $\min{\left(\epsilon_g, \frac{t}{|f^{\prime}(x_0)|}\right)}$ if $f^{\prime}(x_0) \neq 0$, or $\epsilon_g$ when $f^{\prime}(x_0) = 0$.
\end{theorem}

\begin{figure}[ht] \centering
\hspace{-7mm}
\subfigure[{$f(x)=e^x$, $t=0.01$, $\epsilon_g = +\infty$}]
{
\raisebox{-1cm}{\includegraphics[scale=0.23]{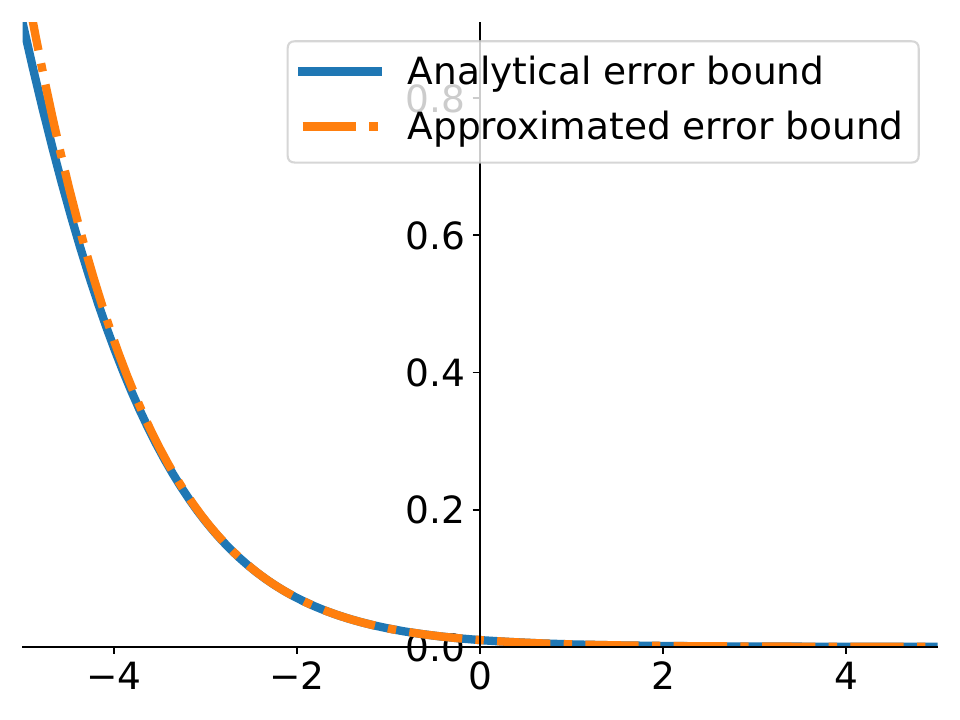}}
}
\hspace{-3mm}
\subfigure[{$f(x)=\ln x$, $t=0.1$, $\epsilon_g = +\infty$}]
{
\raisebox{-1cm}{\includegraphics[scale=0.23]{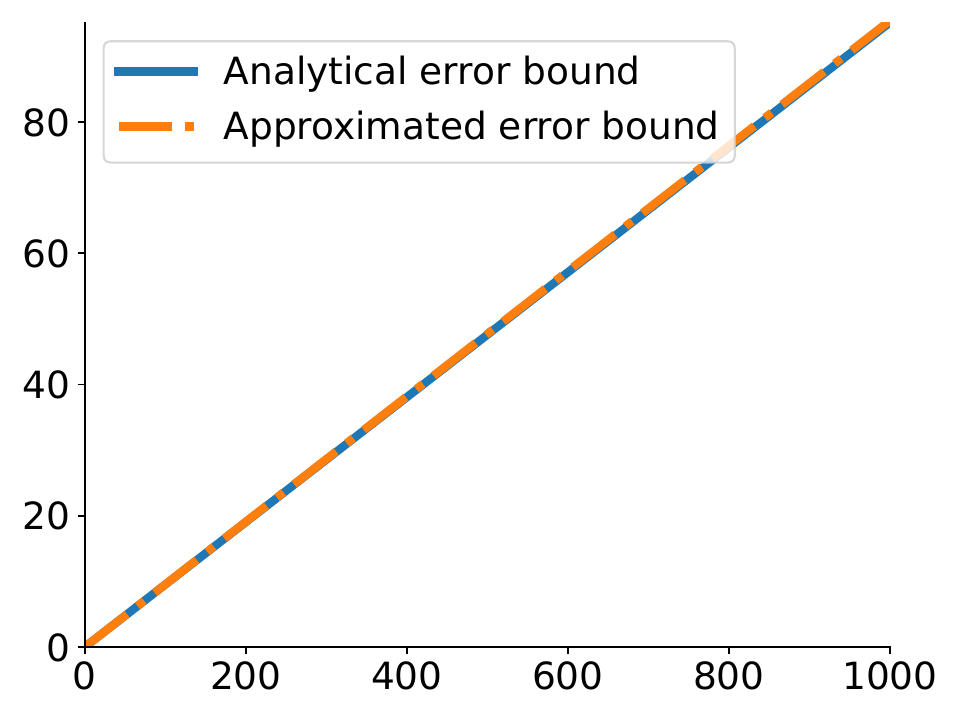}}
}
\hspace{-7mm}

\hspace{-7mm}
\subfigure[{$f(x)=\sin{10x}$, $t=0.1$, $\epsilon_g = +\infty$}]
{
\raisebox{-1cm}{\includegraphics[scale=0.23]{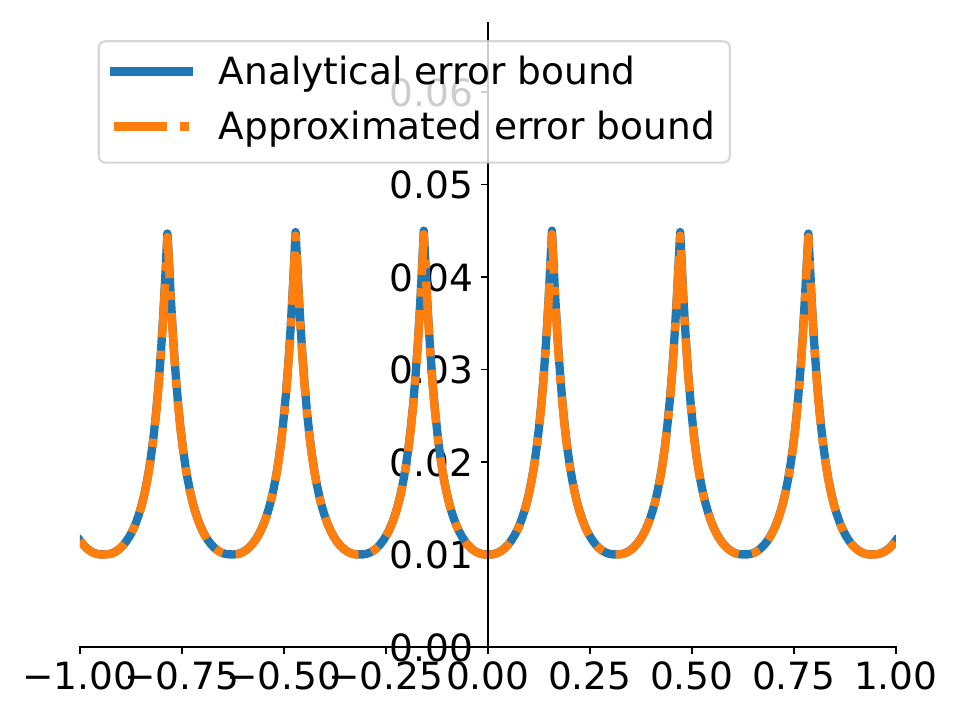}}
}
\hspace{-3mm}
\subfigure[{$f(x)=x^3$, $t=0.01$, $\epsilon_g = +\infty$}]
{
\raisebox{-1cm}{\includegraphics[scale=0.23]{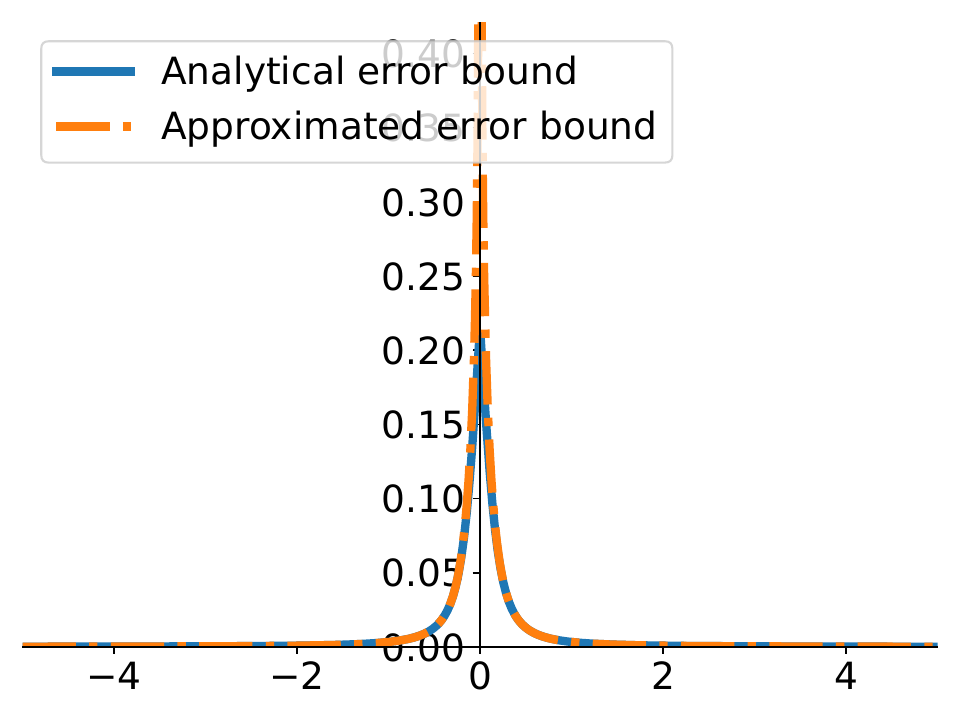}}
}
\hspace{-7mm}
\caption{Analytical and estimation point-wise data error bounds of different QoI function $f(x)$ and error threshold $t$. }
\label{fig:anavsapprox}
\end{figure}

Theorem~\ref{theo:1} and Theorem~\ref{theo:2} cover all the cases for point-wise error-bound estimation.
In Figure~\ref{fig:anavsapprox}, based on different pointwise QoIs and error tolerances, we plot the point-wise data error bounds for different data values from both analytical solutions and our proposed estimation method. 
For various function formats (Figure~\ref{fig:anavsapprox}~(a-c)), QPET acquires accurate error bound estimations compared to the analytical solutions, even when $|f^{\prime\prime\prime}(x)|$ has large values (see Figure~\ref{fig:anavsapprox}~(c) where $f^{\prime\prime\prime}(x) = -1000\cos{10x}$). In Figure~\ref{fig:anavsapprox}~(d), we show an example of $f(x)=x^3$, which has close-to-zero first- and second-order derivatives when $x$ is close to zero. While QPET may produce over-estimated error bounds in this case, the over-estimated error bounds will be cropped by a pre-given or auto-tuned global error bound (to be detailed in Section~\ref{sec:globalt}) and thus not harm the compression process (notice that in Figure~\ref{fig:anavsapprox} we set $\epsilon_g$ as $+\infty$ to focus on the estimation quality from QoI error bounds). As long as QPET presents good estimations for small error bounds (it does), it will be optimized for the compression ratio.

\begin{algorithm}[t]
\caption{\textsc{Point-wise Error bound Estimation with univariate QoI}} \label{alg:point-wise} \footnotesize
\renewcommand{\algorithmiccomment}[1]{/*#1*/}
\begin{flushleft}
\textbf{Input}: input data $X = \left\{x_i\right\}$ of size $n$,  global error bound $\epsilon_g$, QoI function $f$, point-wise QoI error threshold $t$ \\
\textbf{Output}: Approximated optimized point-wise data error bounds $\left\{\epsilon_i\right\}$
\end{flushleft}
\begin{algorithmic} [1]
\STATE $d_1(x) \gets f^{\prime}(x)$ \COMMENT{Get the efficient process for the first-order derivative}
\STATE $d_2(x) \gets f^{\prime\prime}(x)$ \COMMENT{Get the efficient process for the second-order derivative}
\FOR{$i = 1 \to n $}
    \STATE $a \gets d_1\left(x_i\right)$ 
    \STATE $b \gets d_2\left(x_i\right)$ 
    \IF {$b \neq 0$}
    \STATE $\epsilon_i \gets \frac{\sqrt{|a|^2+2|b|t}-|a|}{|b|}$ \COMMENT{Compute $\epsilon_i$ with both derivatives}
    \ELSIF {$a \neq 0$}
    \STATE $\epsilon_i \gets \frac{t}{|a|}$ \COMMENT{Compute $\epsilon_i$ with only first-order derivative}
    \ELSE 
    \STATE $\epsilon_i \gets \epsilon_g$ \COMMENT{set $\epsilon_i$ as $\epsilon_g$}
    \ENDIF 
    \STATE $\epsilon_i \gets \min (\epsilon_g, \epsilon_i)$
\ENDFOR
\RETURN $\left\{\epsilon_i\right\}$
\end{algorithmic}
\end{algorithm}

\subsection{Computing Point-wise Error Bounds with Multivariate QoI}
\label{sec:mulvar}

For multivariate QoIs, we actually deal with QoI functions in the format of $F\left(x_1,\dots,x_n\right)$. $\left\{x_i\right\}$ is a set of data points from the input, which can be a data block, scalar components of a vector, and so on. Given $F\left(x_1,\dots,x_n\right)$, and a tolerance $T$ for which we requires that $\left| F\left(x_1^{\prime},\dots,x_n^{\prime}\right)-F\left(x_1,\dots,x_n\right)\right| \leq t$ ($\left\{x_i^{\prime}\right\}$ are decompressed values of $\left\{x_i\right\}$), we will determine a set of error bounds $\left\{\epsilon_i\right\}$ so that 

$\forall i, \left|x_i^{\prime}-x_i\right| \leq \epsilon_i \implies \left|F\left(x_1^{\prime},\dots,x_n^{\prime}\right)-F\left(x_1,\dots,x_n\right)\right| \leq T $.

In other words, compressing each $x_i$ with an error bound $\epsilon_i$ will lead the QoI error to be constrained. In the following, we first discuss optimizing $\left\{\epsilon_i\right\}$ when $F\left(x_1,\dots,x_n\right)$ is the linear combination of univariate terms, i.e., the variables are separated. Then, we will demonstrate how to extend the methodology to more general cases.  

\subsubsection{Multivariate QoI in variable-separated format}
\label{sec:varsep}
We first demonstrate our strategy for multivariate QoI functions in the format of 
\begin{equation}
\label{eq:linear}
\begin{array}{l}
  F\left(x_1,\dots,x_n\right) = C+\sum_i{ \alpha_i f\left(x_i\right) }
\end{array}
\end{equation}
in which $C$ and $\alpha_i$ are constant coefficients,
and $f$ can be regarded as a point-wise QoI function. Many common multivariate QoIs \cite{jiao2022toward} are in this format, such as region sum, the (weighted) average of a data block, or the difference-based Laplacian operator \cite{su2022understanding}. 
Therefore, it is critical that we present a good solution for this variable-separated format. Suppose we have a set of optimized point-wise tolerances $\left\{t_i\right\}$ from the following optimization problem:

\begin{equation}
\label{eq:linopt}
\begin{split}
&Optimize \left\{t_i\right\} \\
s.t. \ &\forall i, \ \left|f\left(x_i^{\prime}\right)-f\left(x_i\right) \right| \leq t_i \\
&\implies \left|F\left(x_1^{\prime},\dots,x_n^{\prime}\right)-F\left(x_1,\dots,x_n\right)\right| \leq T \\
\end{split}
\end{equation}

Then we can optimize the point-wise data error bound $\left\{\epsilon_i\right\}$ for set $\left\{x_i\right\}$ from those $\left\{t_i\right\}$ with Algorithm~\ref{alg:point-wise}, as in this case we are just bounding point-wise QoI $f$ error within $t_i$. Therefore, by solving Eq.~\ref{eq:linopt} we can find a way to optimize $\left\{\epsilon_i\right\}$. Regarding the optimization target of Eq.~\ref{eq:linopt}, intuitively, we can maximize $\sum_i t_i$. However, this optimization target sometimes leads to skewed distributions of $t_i$ and $\epsilon_i$ values over different $i$. Preliminary works \cite{jiao2022toward} and our experiments have shown that point-wise error bounds with skewed distributions would cause the compression ratio to decrease significantly. Therefore, we select to maximize $\min t_i$, equivalent to ensure $t_i = t$ for each i. According to our preliminary experiments, this scheme outperforms other intuitive strategies, such as aligning $\alpha_i t_i$.
Moreover, if we denote $D_i$ as $f\left(x_i^{\prime}\right)-f\left(x_i\right)$, we will notice that $F\left(x_1^{\prime},\dots,x_n^{\prime}\right)-F\left(x_1,\dots,x_n\right) = \sum_i{\alpha_iD_i}$. To this end, we can convert Eq.~\ref{eq:linopt} to a more concise format:

\begin{equation}
\label{eq:linopt2}
\begin{split}
&Optimize \ t \\
s.t. \ &\forall i, \ |D_i| \leq t \implies \left|\sum_i{\alpha_iD_i}\right| \leq T \\
\end{split}
\end{equation}

Next, we will show two solutions of Eq.\ref{eq:linopt2}, which contribute collaboratively to optimizing $t$, $\left\{\epsilon_i\right\}$, and the compression ratio. First, we have a solution of $t$ that deterministically matches Eq.\ref{eq:linopt2}:

\begin{theorem}
    \label{theo:3}
    If $t = \frac{T}{\sum_i\left|\alpha_i\right|}$, then $ \ |D_i| \leq t \implies \left|\sum_i{\alpha_iD_i}\right| \leq T$.
\end{theorem}

\begin{proof}
    $\left|\sum_i{\alpha_iD_i}\right| \leq \sum_i{\left|\alpha_i\right|\left|D_i\right|} \leq t\sum_i{\left|\alpha_i\right|} = T$.
\end{proof}

Unfortunately, Theorem~\ref{theo:3} sometimes presents over-conservative values for $t$. Fortunately, we have another strategy for determining $t$. If we regard the $\left\{D_i\right\}$ as random variables with the following assumptions: 1) $\left\{D_i\right\}$ are independent; 2) $D_i$ has symmetric distribution regarding zero (so $E\left(D_i\right) = 0$). Since $|D_i| \leq t$, according to \cite{hdp,hds}, $\left\{D_i\right\}$ are sub-Gaussian, and the following concentration inequality holds for each $T > 0$ \cite{hoeffding,hdp,hds}:

\begin{equation}
\label{eq:hoeffdin}
\begin{array}{l}
  P\left( \left|\sum_i{\alpha_iD_i}\right| \geq T \right) \leq 2e^{-\frac{T^2}{2\sum_i{\alpha_i^2\sigma_i^2}}}
\end{array}
\end{equation}

$\sigma_i^2$ is called the \textit{variance proxy} of $D_i$. For finite-bounded $|D_i| \leq t$, we have $\sigma_i \leq t$ \cite{hdp,hds}. QPET uses a single estimation $\sigma_0 = \frac{t}{c}$ for all $D_i$, in which $c \leq 1$ and $\sigma_0$ represents the estimated standard variance of $D_i$. For example, $c = \sqrt{3}$ and $\sigma_0 = \frac{t}{\sqrt{3}}$ correspond to the uniform distribution over $\left[-t,t\right]$. With $\sigma_0$ and Eq.~\ref{eq:hoeffdin}, we can set up another estimation of an optimized $t$ for Eq.~\ref{eq:linopt2}:

\begin{theorem}
    \label{theo:4}
    A multivariate QoI function $F\left(x_1,\dots,x_n\right) = C+\sum_i{\alpha_i f\left(x_i\right)}$ and the corresponding QoI error threshold $T$ are given. On $\left\{x_1,\dots,x_n\right\}$, if the point-wise QoI error $D_i$ is $\sigma_i$-Sub-Gaussian in which $\sigma_i \leq \frac{\max D_i}{c}$, for each confidence level of $\beta$, taking point-wise QoI error threshold $t = \max\left|f\left(x_i^{\prime}\right)-f\left(x_i\right)\right| = cT\sqrt{\frac{1}{2\sum_i{\alpha_i^2}\ln{\frac{2}{1-\beta}}}}$, the global QoI error threshold can be guaranteed in a confidence level of $\beta$, i.e.: $P\left(\left|F\left(x_1^{\prime},\dots,x_n^{\prime}\right)-F\left(x_1,\dots,x_n\right)\right| \leq T \right) \geq \beta$.
\end{theorem}

\begin{proof}
    Using $\sigma_i = \frac{t}{c}$, for $\left\{x_1,\dots,x_n\right\}$, from Eq. \ref{eq:hoeffdin} we have  
    
    $P\left(\left|F\left(x_1^{\prime},\dots,x_n^{\prime}\right)-F\left(x_1,\dots,x_n\right)\right| \geq T \right) = 
    P\left( \left|\sum_i{\alpha_iD_i}\right| \geq T \right)$ 
    
    $\leq 2e^{-\frac{c^2T^2}{2t^2\sum_i{\alpha_i^2}}} = 2e^{-\ln{\frac{2}{1-\beta}}} = 1-\beta$

    Therefore, $P\left(\left|F\left(x_1^{\prime},\dots,x_n^{\prime}\right)-F\left(x_1,\dots,x_n\right)\right| \leq T \right)$
    
    $= 1- P\left(\left|F\left(x_1^{\prime},\dots,x_n^{\prime}\right)-F\left(x_1,\dots,x_n\right)\right| \geq T \right) \geq \beta$.
\end{proof}

Because Theorem~\ref{theo:3} and ~\ref{theo:4} fits in different cases when optimizing $t$ in Eq.~\ref{eq:linopt2}. In practice, we select the larger one between those two estimations to optimize the compression ratio. For a small variable number $n$, Theorem~\ref{theo:3} can often provide better estimations and vice versa. Table~\ref{tab:t34} shows a few examples of the point-wise QoI ($f(x)$) error tolerances deduced from Theorem~\ref{theo:3} and \ref{theo:4}. As the number of variables increases, Theorem~\ref{theo:4} provides more aggressive estimations.

Admittedly, Theorem~\ref{theo:4} needs certain preliminary conditions of $\left\{D_i\right\}$, which are not always true. 
The variance of compression errors is dependent on diverse factors, such as data characteristics and input error bound, and $\left\{D_i\right\}$ are not theoretically independent of each other (such as in interpolation-based data compression). Nevertheless, existing research~\cite{szinterp,qoz,HPEZ} showed that, in practical cases, distributions of $\left\{D_i\right\}$ are not severely distorted from the assumptions, exhibiting distributions that are close to uniform or Gaussian, and the autocorrelation of decompression errors drop rapidly as the compression accuracy increases \cite{qoz}. Moreover, to handle the unclear characteristics of $\left\{D_i\right\}$, Theorem~\ref{theo:4} can still acquire adequate QoI error threshold estimations by selecting a conservative value of $c$. 
To optimize the compression ratio, Theorem~\ref{theo:4} does not bring a strict guarantee for bounding all QoI errors. In Section~\ref{sec:qoival}, we will discuss how QPET strictly guarantees the QoI error threshold and can optimize the compression ratio through more aggressive data error-bound settings.

\begin{table}[]
\centering
\caption{Point-wise QoI ($f(x)$) error thresholds from different multivariate QoI function $F$ and variable numbers ($c=2$, $\beta=0.9999$). Error threshold of $F(X)$ is $T$.}
\label{tab:t34}
\footnotesize
\renewcommand{\arraystretch}{1.2} 
\resizebox{0.99\columnwidth}{!}{%
\begin{tabular}{|c|c|c|c|}
\hline
\textbf{QoI} & \textbf{Variable Num. ($n$)} & \textbf{$t$ from Theo.~\ref{theo:3}}& \textbf{$t$ from Theo.~\ref{theo:4}}\\ \hline
$f(x)+f(y)+f(z)$ &    3    &     \textbf{$0.33T$ }     &   $0.26T$    \\ \hline
\multirow{2}{*}{$\frac{1}{n}\sum f(x)$} &     $2^3$   &      $T$     &     \textbf{$1.27T$}   \\ \cline{2-4}
 &    $4^3$     &    $T$        &    \textbf{$3.6T$}   \\ \hline
\end{tabular}
}
\end{table}

\subsubsection{Multivariate QoI in non-variable-separated format}

 Beyond the variable-separated format described in Eq.~\ref{eq:linear}, regarding a non-variable-separated multivariate QoI function $F\left(x_1,\dots,x_n\right)$, to separate the variables so that techniques described in Section~\ref{sec:varsep} can be applied, we use its approximation for computing point-wise error bounds, leveraging the first-order differentials of $F$ (using higher-order differentials will bring cross partial derivatives, failing to separate the variables). Specifically, we have: 

\begin{equation}
\label{eq:linearapprox}
\begin{array}{l}
F\left(x_1^{\prime},\dots,x_n^{\prime}\right) \approx F\left(x_1,\dots,x_n\right)+\sum_i{ \frac{\partial F}{\partial x_i} \left(x_i\right)\left(x_i^{\prime}-x_i\right) } 
\end{array}
\end{equation}

Eq.~\ref{eq:linearapprox} falls into the format of Eq.~\ref{eq:linear}, with $C = F\left(x_1,\dots,x_n\right)$, $\alpha_i = \frac{\partial F}{\partial x_i}\left(x_i\right)$, and $f\left(x\right) = x-x_i$ (equivalent to $f(x)=x$). Therefore, we can use Eq.~\ref{eq:linearapprox} and Theorem~\ref{theo:3}/\ref{theo:4} to determine the point-wise error bounds from $F$. It is worth noticing that, in this case, for different $\left\{x_i\right\}$, we will have different $t$ from Theorem~\ref{theo:3} and ~\ref{theo:4} as there are different $\alpha_i$ values from the partial derivatives.

In Algorithm~\ref{alg:regional}, we combine the computation of point-wise data error bounds from multivariate QoI in either linear or non-linear format. When dealing with the linear format, the computation can often be simplified. For example, for linear multivariate QoIs, the $t$ in Line 14 can be shared over data regions.

 \begin{algorithm}[t]
\caption{\textsc{Point-wise Error bound Estimation with multi-variate QoI Constraint}}
\label{alg:regional} \footnotesize
\renewcommand{\algorithmiccomment}[1]{/*#1*/}
\begin{flushleft}
\textbf{Input}: input data $X = \left\{x_i\right\}$ , Data regions $\mathcal{R} = \left\{R_i\right\}$ ( $R_i=\left\{r_{ij}\right\}$, $id(r_{ij})$ is the global index of $r_{ij}$ so $r_{ij} = x_{id(r_{ij})}$),  global error bound $\epsilon_g$, multivariate QoI function $F$, QoI error threshold $T$, estimation parameter $c$, and confidence level $\beta$. \\
\textbf{Output}: Point-wise data error bounds $\left\{\epsilon_i\right\}$ For multivariate QoI error constraint.
\end{flushleft}
\begin{algorithmic} [1]
\FOR{$i = 1 \to |X| $}
\STATE    $\epsilon_i \gets \epsilon_g$ \COMMENT{Initialize all $\epsilon_i$ with global error bound}
\ENDFOR
\FOR[Iterate through regions]{$R_i \in \mathcal{R} $  } 
    \IF [Linear Format]{$F\left(R_i\right) = C +\sum_j{ a_j g\left(r_{ij}\right)}$}
    \STATE $f \gets g $
    \STATE $\left\{\alpha_j\right\} \gets \left\{a_j\right\}$
    \ELSE [Non-linear]
    \STATE $f \gets f(x)= x $
    \STATE $\left\{\alpha_j\right\} \gets \left\{\frac{\partial F}{\partial x_{j}}(r_{ij})\right\}$  
    \ENDIF 
    \STATE $t_1 \gets \frac{T}{\sum_i\left|\alpha_i\right|}$ \COMMENT{Theorem~\ref{theo:3}}
    
    \STATE $t_2 \gets cT\sqrt{\frac{1}{2\sum_j{\alpha_j^2}\ln{\frac{2}{1-\beta}}}}$ \COMMENT{Theorem~\ref{theo:4}}
    \STATE $t \gets \max (t_1, t_2)$ 
    \FOR [Iterate through data value in regions ]{ $r_{ij} \in R_i$ } 
    \STATE $\epsilon \gets \operatorname{Compute\_eb\_univar}(r_{ij}, t, f, \epsilon)$ \COMMENT{By Algorithm~\ref{alg:point-wise}}
    \STATE $\epsilon_{id(r_{ij})} \gets \min \left(\epsilon_{id(r_{ij})}, \epsilon \right)$ \COMMENT{update $\epsilon_{id(r_{ij})}$}
    \ENDFOR
    
\ENDFOR
\RETURN $\left\{\epsilon_i\right\}$
\end{algorithmic}
\end{algorithm}

\subsection{Optimizing Global Error Bound}
\label{sec:globalt}
After QPET acquires the point-wise error bound for each data point $x_i$, gathering up a set of error bounds $\left\{\epsilon_i\right\}$, QPET auto-tunes a new global error bound $\epsilon_g$, cropping all larger point-wise error bounds to $\epsilon_g$ in the compression. It is for two purposes: 1) For compressors that support point-wise accuracy (like SZ3/HPEZ), QPET will store the quantized point-wise error bound, and cropping most error bounds will significantly reduce their storage, which improves the overall compression ratio in turn; 2) For compressors that do not support point-wise accuracy (like SPERR), QPET needs an optimized global accuracy setting as the compression configuration.

To this end, we proposed a two-step dynamic global error-bound auto-tuning strategy in QPET. 
After acquiring the set of point-wise error bounds $\left\{\epsilon_i\right\}$, in the first step, a set of quantiles (20\%, 10\%, 5\%, 2\%, 1\%, 0.5\%, 0.25\%) is drawn from it as global error bound candidates, and then QPET identify the best from them in a sample-and-tests scheme described in \cite{qoz}.
In the second step, QPET further reduces $\epsilon_g$ over the distribution of $\left\{\epsilon_i\right\}$ until a steep decrease is detected. According to our preliminary experiments, saving fewer error-bound values is often more beneficial in improving the compression ratio than applying a slightly larger global error bound. 

The details of the described strategy are featured in Algorithm~\ref{alg:globaltune}. In Lines 1 to 7, QPET optimizes the global error bound by tests on sampled data. In Line 13, QPET checks whether the current error bound $\epsilon$ is still over a gentle linear slope between the initial $\epsilon_0$ and a termination value $\epsilon_t$ (e.g., $\epsilon_t = 0.95\epsilon_0$), and stop the decreasing of $\epsilon_g$ if it will beneath that slope.
The $\operatorname{KthSmallest}()$ operations in Algorithm~\ref{alg:globaltune} can be implemented by fast-selection and only need to be performed on reduced ranges of $\left\{\epsilon_i\right\}$. Therefore, all the operations in Algorithm~\ref{alg:globaltune} can be efficiently done with $\mathcal{O}(n+qn\log qn)$ time complexity in average, in which $q$ is the quantile of $\epsilon_0$. In practice, QPET skips the second part of tuning (from Line 8) when $q$ is larger than a threshold (0.005) to guarantee efficiency, bringing negligible overhead to the overall compression pipeline.

\begin{algorithm}[t]
\caption{\textsc{Global error bound auto-tuning}} \label{alg:globaltune} \footnotesize
\renewcommand{\algorithmiccomment}[1]{/*#1*/}
\begin{flushleft}
\textbf{Input}: point-wise data error bounds $\left\{\epsilon_i\right\}$ with size $n$, a dropping threshold coefficient $c_0 \leq 1$ \\
\textbf{Output}: Auto-tuned global error bound $\epsilon_g$
\end{flushleft}
\begin{algorithmic} [1]
\STATE    $E_c \gets \{\}$
\FOR {$q \in \{0.2, 0.1, 0.05, 0.02, 0.01, 0.005, 0.0025\}$}
\STATE $k \gets \left \lfloor qn \right \rfloor$ 
\STATE $\epsilon \gets \operatorname{KthSmallest}\left(\left\{\epsilon_i\right\}, k\right)$ \COMMENT{$\epsilon$ is the k\textsuperscript{th}-smallest of $\left\{\epsilon_i\right\}$}
\STATE $E_c.\operatorname{insert}((\epsilon,k))$
\ENDFOR   
\STATE $ \epsilon_0, k_0 \gets \operatorname{cmpTest\_BestCR}\left(E_c\right)$ \COMMENT{$\epsilon_g$ achieves the best compression ratio in compression tests among all error bounds in $E_c$, save its current value in $\epsilon_0$}
\STATE $\epsilon_g \gets \epsilon_0$
\STATE $E_p \gets \operatorname{partialSort}\left(\left\{\epsilon_i\right\}, k_0\right)$ \COMMENT{$E_p$ is the sorted $k$ smallest values in $\left\{\epsilon_i\right\}$}
\STATE $k \gets k_0-1 $ \COMMENT{Start checking smaller error bounds}
\WHILE {$k \geq 0$ \COMMENT{we regard minimum as 0\textsuperscript{th}-smallest}}
\STATE $\epsilon \gets E_p[k] $
\IF {$\epsilon \geq (c_0+\frac{k}{k_0}(1-c_0))\epsilon_0$ \COMMENT{$\epsilon$ is above a slope}}
\STATE $\epsilon_g \gets \epsilon$ \COMMENT{Update $\epsilon_g$}
\ELSE 
\STATE \textbf{break} \COMMENT{Stop decreasing $\epsilon_g$}
\ENDIF 
\STATE $k \gets k-1 $
\ENDWHILE
\RETURN $\epsilon_g$
\end{algorithmic}
\end{algorithm}

\subsection{QoI Error Validation and Correction}
\label{sec:qoival}
As previously described in Section~\ref{sec:problem}, the design purpose of QPET is to optimize the compression ratio when strictly constraining the QoI error threshold. To fulfill this target, the point-wise and global error-bound calculations demonstrated in Section~\ref{sec:univar}, Section~\ref{sec:mulvar}, and Section~\ref{sec:globalt} do not conservatively make the QoI value strictly preserved on each data element, but require a subsequent process to correct and store the out-of-bound data points (i.e., outliers). By selecting proper error-bound values, QPET can significantly reduce the average storage cost of each compressed data value, meanwhile introducing negligible overheads for correcting outliers as they are only a tiny portion ($<1\%$) of the whole input data. Specifically, QPET detects and corrects the outliers in the following scheme: Given the decompressed data (which are naturally acquired after the compression process), QPET computes the QoI values on it and compares them to the original QoI values. For out-of-bound univariate QoI errors, QPET losslessly encodes and stores all the corresponding original data values. For each out-of-bound multivariate QoI error, there are several related data values. To minimize the number of data values that are losslessly stored, QPET stores those values one by one and updates the QoI result after each value update until the QoI error gets bounded.

\section{Evaluations}
\label{sec:evaluation}

We evaluate QPET using six real-world datasets and diverse QoIs. 
In particular, we integrate QPET into three top-performing compressors (SZ3~\cite{sz3}, HPEZ~\cite{HPEZ}, and SPERR~\cite{SPERR}) and compare them with diverse baselines in terms of QoI preservation.

\subsection{Experimental Setup}
\subsubsection{Experimental environment and datasets}

We perform the evaluations on 6 real-world scientific datasets from diverse domains (details in Table~\ref{tab:dataset information}). Experiments are operated on the Purdue Anvil computing cluster \cite{Cluster-Anvil} (each node is equipped with two 64-core AMD EPYC 7763 CPUs and 512GB DDR4 memory).

\begin{table}[h]
    \centering
    \caption{Information of the datasets in experiments}  
    \footnotesize
\resizebox{0.99\columnwidth}{!}{  
    \begin{tabular}{|c|c|c|c|c|}
    \hline
    App.&\# fields& Dimensions & Total Size& Domain\\
    \hline
    Miranda & 7 & 256$\times$384$\times$384& 1GB& Turbulence \\
    \hline
    Hurricane & 13 & 100$\times$500$\times$500 &1.2GB& Weather\\
    \hline
    RTM &11&449$\times$449$\times$235&2.0GB&Seismic Wave\\
    \hline
    NYX & 6 & 512$\times$512$\times$512 &3.1GB& Cosmology\\
    \hline
    SEGSalt &3&1008$\times$1008$\times$352&4.0GB&Geology\\
    \hline

    SCALE-LetKF & 12 & 98$\times$1200$\times$1200 & 6.3GB&Climate\\
    \hline

    \end{tabular}}
    \label{tab:dataset information}
\end{table}

 \subsubsection{Baselines}
Besides the QPET-integrated compressors, we included several existing solutions for QoI-preserving scientific lossy compression in the evaluations, which are in 2 categories: 1) \textbf{Parameter-search-based solutions}: general-purpose compressors (SZ3/HPEZ/SPERR) cannot directly bound specified QoI errors, so we apply parameter-search methods on top of them to figure out the best-fit (yielding highest compression ratio) data error bound for preserving the QoI. The parameter-search methods, including binary-search-based and FraZ~\cite{underwood2020fraz}, are from OptZConfig~\cite{underwood2022optzconfig}, the state-of-the-art scientific lossy compression parameter-search toolkit. When applying, we slightly revised those methods to get them better adapted and accelerated in the QoI-preserving tasks. Those baselines are named SZ3/HPEZ/SPERR-OptZ-R (R is short for revised), and they represent the best (fastest) results from both parameter-search methods on every base compressor. 
The parameter-search process often requires multiple iterations of data compression and QoI validation to fit the QoI error bound, making it typically quite slow.
(2) \textbf{Direct QoI-preserving solutions}: \cite{jiao2022toward} provided the SZ3-based QoI-preserving compressor QoI-SZ3, and we further ported its QoI-preserving features to HPEZ, creating QoI-HPEZ. Moreover, we evaluated MGARD-QoI~\cite{mgardqoi}. Those QoI-preserving compressors have limited support for diverse QoI formats. QoI-SZ3/HPEZ only supports square and logarithm (also their block averages), and MGARD-QoI only supports linear QoIs. So, we only tested them on the QoIs supported. Other existing QoI-integrated compressors are designed for different tasks (e.g., cpSZ~\cite{liang2018efficient} only works for critical points in vector field data), so they are not included in our evaluation baselines.

\subsubsection{QoI functions, experimental configurations, and evaluation metrics}

\label{sec:config}

Table~\ref{tab:qoilist} shows the QoI functions in the evaluation tasks. Among them, there are three different categories: point-wise, regional, and vector. They have diverse mathematical formats, and for many among them (such as $\tanh{x}$, $\frac{1}{n}\sum{x^3}$, and vector QoIs), QPET is the first framework that supports compression while preserving those QoIs. The selection of QoI functions in our evaluation is based on existing investigations and analysis \cite{jiao2022toward,su2022understanding,wu2024error} of QoIs in practical scientific data analysis tasks, such as physical transform (kinetic energy as velocity's square), and clustering ($\sqrt{x^2+y^2+z^2}$ is the distance from origin when $x$, $y$, and $z$ are coordinates).
\begin{table}[ht]
\centering
\caption{QoI functions in the evaluation} 
\label{tab:qoilist}
\footnotesize
\renewcommand{\arraystretch}{1.3} 
\begin{tabular}{|c|c|c|c|}
\hline

\textbf{QoI type}                   & \textbf{QoI function}      & \textbf{QoI type}                  & \textbf{QoI function}                                              \\ \hline
\multirow{5}{*}{\textbf{Pointwise}} & $x^2$         & \multirow{3}{*}{\textbf{Regional}} & $x$ (average)                                      \\ \cline{2-2} \cline{4-4} 
                                    &     $x^3$                 &                                    & $x^2$ (average)                                                    \\ \cline{2-2} \cline{4-4} 
                                    & $\log_2x$         &                    &   $x^3$ (average)                                     \\ \cline{2-4} 
                                    &$\sin{10x}$  & \multirow{2}{*}{\textbf{Vector}}         &   $x^2+y^2+z^2$     \\\cline{2-2} \cline{4-4} 
                                    &$\tanh{x}$                     &                                    &$\sqrt{x^2+y^2+z^2}$ \\ \hline
\end{tabular}
\end{table}

For the OptZ-R parameter search, we set an early termination condition that triggers when a maximum QoI error between 90\% and 100\% of the required threshold is found, as it presents a near-optimal compression ratio with a reasonable search time.
Regarding the compression configurations, we apply the default optimization level and compression-ratio-preferred mode for HPEZ. Regarding QPET parameters in Algorithms \ref{alg:regional}, we set $c=3$, $\beta = 0.999$ for SPERR, $c=2$, $\beta = 0.999$ for HPEZ, and $c=2$, $\beta=0.99999$ for SZ3. On SZ3 and HPEZ, the autocorrelation of decompression errors are relatively high when the error bound is large \cite{qoz}, so we linearly dynamically decrease $c$ when $\tau$ increases over $10^{-3}$, eventually to 1.0 when $\tau$ becomes $10^{-2}$.

In evaluating the compression performance, the following widely adopted metrics \cite{z-checker,jiao2022toward} are used:  (1) Compression and decompression speeds (throughputs). (2) Compression ratio $\operatorname{CR} = \frac{|X|}{|C|}$, which is the input data size $|X|$ divided by the data size $|C|$; (3) Bit rate $\operatorname{BR} = \frac{|C|*8*\operatorname{sizeof}(x)}{|X|}$, which is the number of bits in compressed data to store each value in the input. (3) Maximum data error and QoI error between the input and output; 

\begin{figure}[ht]
    \centering
\includegraphics[width=0.99\linewidth]{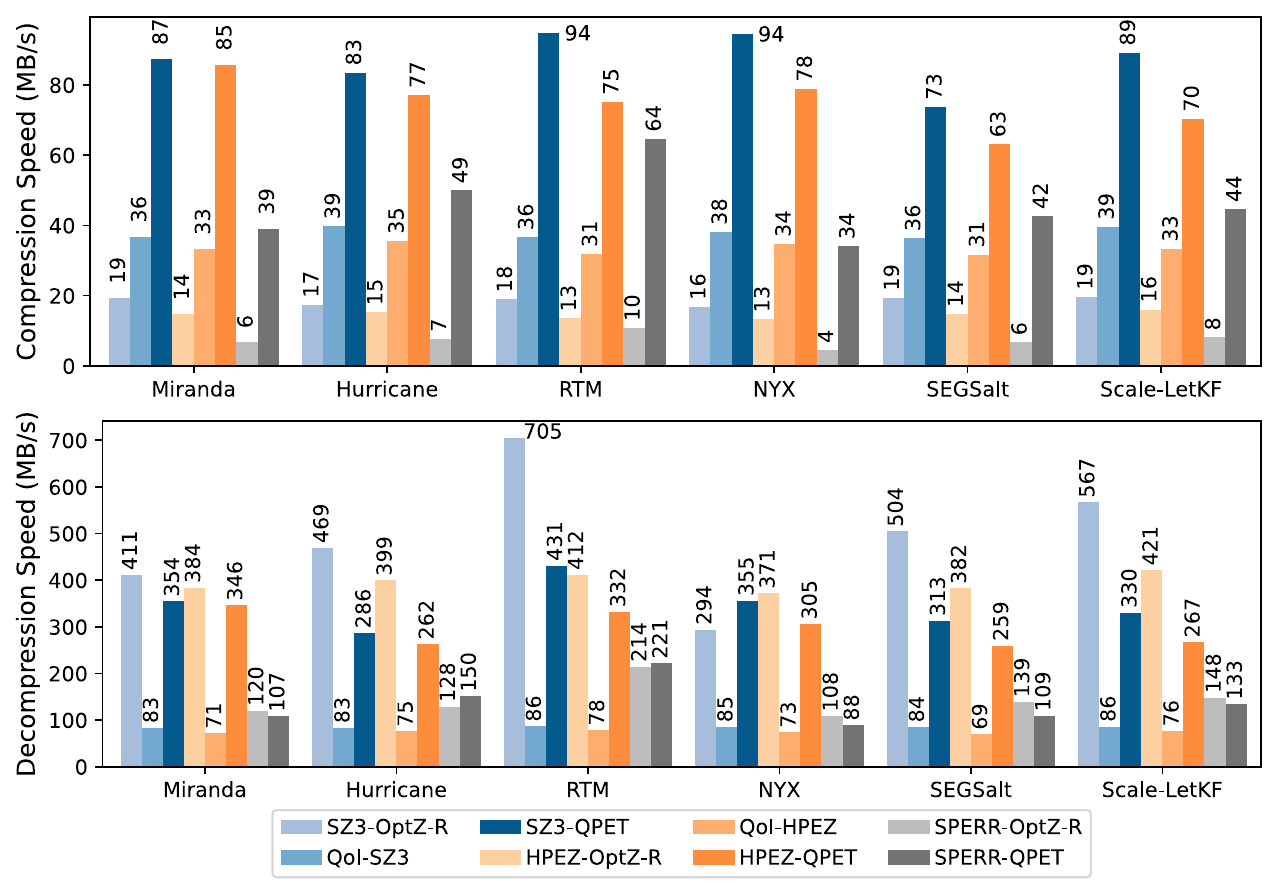}
    \caption{Compression and decompression speed for $Q(x)=x^2$ and $\tau = 1\mathrm{e}{-3}$.}
    \label{fig:speed-x2}
\end{figure}

 \begin{figure}[ht]
    \centering
\includegraphics[width=0.99\linewidth]{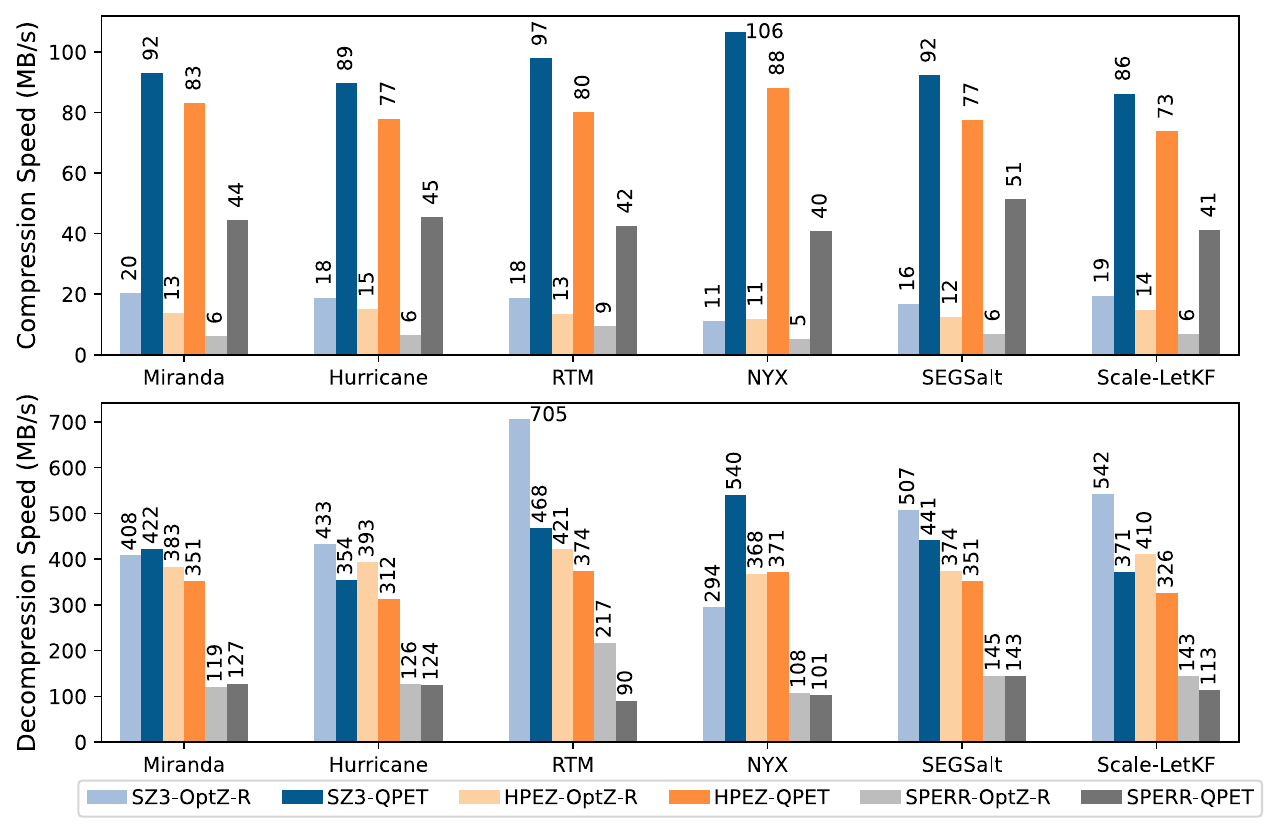}
    \caption{Compression and decompression speed for $Q(x)=x^3$ and $\tau = 1\mathrm{e}{-3}$.}
    \label{fig:speed-x3}
\end{figure}

\newcommand \subfigsize{0.39}

\subsection{Evaluation Results}
Here, we propose our evaluation results for QPET, which prove its advantages and versatility. Due to page limits, we selected the most representative results to be presented in this paper.
\subsubsection{Outperforming speeds}
\label{sec:eva:speeds}

First, we measure and profile the execution speeds of baselines and QPET-integrated compressors to examine how QPET has accelerated the QoI-preserving error-bounded lossy compression. From the compression and decompression speeds on the Anvil cluster with the QoI functions $Q(x)=x^2$ and $Q(x)=x^3$ and the relative QoI error threshold $\tau=1\mathrm{e}{-3}$, reported in Figure~\ref{fig:speed-x2} to \ref{fig:speed-x3} (no results from QoI-SZ3/HPEZ for $Q(x)=x^3$ because their current implementation don't support this QoI), we can conclude as follows: compared to existing parameter-search-based baselines and QoI-preseving compressors, QPET-integrated compressors have significantly improved the compression throughputs (4x $\sim$ 6x of *-OptZ-R and 2x $\sim$ 2.5x of QoI-SZ3/HPEZ), with similar decompression throughput with the fast general-purpose compressors. Parameter search on general-purpose compressors (*-OptZ-R) needs to perform repeated executions for acquiring a certain QoI error threshold, and in our evaluations, this task may cost 4 to 10 or even more compression executions and QoI validations. Moreover, QPET brings quite limited time overheads in the decompression process (because neither error-bound computation nor QoI validation is needed), and sometimes it even improves the decompression speed over baselines due to the reduced accuracy required in the compression to ensure the same QoI error threshold. 
Last, compared to the existing QoI-SZ3/HPEZ, for the same use cases, QPET gains 100\% $\sim$ 150\% compression speed improvement and up to 400\% decompression speed improvement, which is highly attributed to the efficient error-bound selection and tuning process proposed by QPET, and also the revised implementation in storing and recovering the point-wise error bounds.

\subsubsection{Representative showcases}

\begin{table*}[ht]
\scriptsize
\caption{Showcases of QoI-preserving error-bounded lossy compression. For blocked QoIs, $n = 4^3$ (i.e., average on 4x4x4 blocks). $\epsilon$: data error bound. $\tau$: QoI error threshold. \#It: Number of iterations. CR: Compression ratio. $T_c$: Compression throughput (in MB/s). $T_d$: Decompression throughput (in MB/s). All experiments are performed on Purdue Anvil, and all error bounds are relative (absolute value/value range). N/A indicates that the baseline is not available in the case.}
\label{tab:showcase}
\begin{tabular}{|ccc|cccc|cccc|cccc|cccc|cccc|}
\hline
\multicolumn{3}{|c|}{QoI}                                                                                         & \multicolumn{4}{c|}{$Q(x)=x^2$}                                                                            & \multicolumn{4}{c|}{$Q(x)=x^3$}                                                                            & \multicolumn{4}{c|}{$Q(x)=\log_2x$}                                                                        & \multicolumn{4}{c|}{$Q(x)=\sin{10x}$}                                                                            & \multicolumn{4}{c|}{$Q(x)=\tanh{x}$}                                                                           \\ \hline
\multicolumn{3}{|c|}{Data field}                                                                                  & \multicolumn{4}{c|}{SegSalt-Pressure2000}                                                                  & \multicolumn{4}{c|}{RTM-3500}                                                                              & \multicolumn{4}{c|}{NYX-Baryon Density}                                                                    & \multicolumn{4}{c|}{Miranda-Pressure}                                                                            & \multicolumn{4}{c|}{Scale-LetKF-RH}                                                                            \\ \hline
\multicolumn{1}{|c|}{$\epsilon$}                 & \multicolumn{1}{c|}{$\tau$}                     & Compressor   & \multicolumn{1}{c|}{\#It} & \multicolumn{1}{c|}{CR}            & \multicolumn{1}{c|}{$T_c$}        & $T_d$ & \multicolumn{1}{c|}{\#It} & \multicolumn{1}{c|}{CR}            & \multicolumn{1}{c|}{$T_c$}        & $T_d$ & \multicolumn{1}{c|}{\#It} & \multicolumn{1}{c|}{CR}            & \multicolumn{1}{c|}{$T_c$}        & $T_d$ & \multicolumn{1}{c|}{\#It} & \multicolumn{1}{c|}{CR}            & \multicolumn{1}{c|}{$T_c$}       & $T_d$        & \multicolumn{1}{c|}{\#It}      & \multicolumn{1}{c|}{CR}            & \multicolumn{1}{c|}{$T_c$}       & $T_d$ \\ \hline
\multicolumn{1}{|c|}{\multirow{8}{*}{$10^{-1}$}} & \multicolumn{1}{c|}{\multirow{8}{*}{$10^{-2}$}} & SZ3-OptZ-R   & \multicolumn{1}{c|}{9}    & \multicolumn{1}{c|}{283}           & \multicolumn{1}{c|}{21}           & 693   & \multicolumn{1}{c|}{9}    & \multicolumn{1}{c|}{167}           & \multicolumn{1}{c|}{19}           & 695   & \multicolumn{1}{c|}{23}   & \multicolumn{1}{c|}{14.1}          & \multicolumn{1}{c|}{3}            & 226   & \multicolumn{1}{c|}{12}   & \multicolumn{1}{c|}{101}           & \multicolumn{1}{c|}{7}           & 379          & \multicolumn{1}{c|}{12}        & \multicolumn{1}{c|}{9.88}          & \multicolumn{1}{c|}{8}           & 182   \\ \cline{3-23} 
\multicolumn{1}{|c|}{}                           & \multicolumn{1}{c|}{}                           & HPEZ-OptZ-R  & \multicolumn{1}{c|}{9}    & \multicolumn{1}{c|}{448}           & \multicolumn{1}{c|}{14}           & 413   & \multicolumn{1}{c|}{6}    & \multicolumn{1}{c|}{231}           & \multicolumn{1}{c|}{20}           & 428   & \multicolumn{1}{c|}{3}    & \multicolumn{1}{c|}{14.7}          & \multicolumn{1}{c|}{21}           & 182   & \multicolumn{1}{c|}{12}   & \multicolumn{1}{c|}{151}           & \multicolumn{1}{c|}{6}           & 369          & \multicolumn{1}{c|}{12}        & \multicolumn{1}{c|}{8.78}          & \multicolumn{1}{c|}{8}           & 214   \\ \cline{3-23} 
\multicolumn{1}{|c|}{}                           & \multicolumn{1}{c|}{}                           & SPERR-OptZ-R & \multicolumn{1}{c|}{6}    & \multicolumn{1}{c|}{413}           & \multicolumn{1}{c|}{10}           & 144   & \multicolumn{1}{c|}{9}    & \multicolumn{1}{c|}{237}           & \multicolumn{1}{c|}{10}           & 220   & \multicolumn{1}{c|}{23}   & \multicolumn{1}{c|}{19.4}          & \multicolumn{1}{c|}{2}            & 71    & \multicolumn{1}{c|}{12}   & \multicolumn{1}{c|}{132}           & \multicolumn{1}{c|}{3}           & 118          & \multicolumn{1}{c|}{12}        & \multicolumn{1}{c|}{11.4}          & \multicolumn{1}{c|}{4}           & 60    \\ \cline{3-23} 
\multicolumn{1}{|c|}{}                           & \multicolumn{1}{c|}{}                           & QoI-SZ3      & \multicolumn{1}{c|}{}     & \multicolumn{1}{c|}{458}           & \multicolumn{1}{c|}{36}           & 88    & \multicolumn{1}{c|}{}     & \multicolumn{1}{c|}{N/A}           & \multicolumn{1}{c|}{N/A}          & N/A   & \multicolumn{1}{c|}{}     & \multicolumn{1}{c|}{21.9}          & \multicolumn{1}{c|}{19}           & 80    & \multicolumn{1}{c|}{}     & \multicolumn{1}{c|}{N/A}           & \multicolumn{1}{c|}{N/A}         & N/A          & \multicolumn{1}{c|}{}          & \multicolumn{1}{c|}{N/A}           & \multicolumn{1}{c|}{N/A}         & N/A   \\ \cline{3-23} 
\multicolumn{1}{|c|}{}                           & \multicolumn{1}{c|}{}                           & QoI-HPEZ     & \multicolumn{1}{c|}{}     & \multicolumn{1}{c|}{687}           & \multicolumn{1}{c|}{30}           & 70    & \multicolumn{1}{c|}{}     & \multicolumn{1}{c|}{N/A}           & \multicolumn{1}{c|}{N/A}          & N/A   & \multicolumn{1}{c|}{}     & \multicolumn{1}{c|}{24}            & \multicolumn{1}{c|}{17}           & 71    & \multicolumn{1}{c|}{}     & \multicolumn{1}{c|}{N/A}           & \multicolumn{1}{c|}{N/A}         & N/A          & \multicolumn{1}{c|}{}          & \multicolumn{1}{c|}{N/A}           & \multicolumn{1}{c|}{N/A}         & N/A   \\ \cline{3-23} 
\multicolumn{1}{|c|}{}                           & \multicolumn{1}{c|}{}                           & SZ3-QPET     & \multicolumn{1}{c|}{}     & \multicolumn{1}{c|}{689}           & \multicolumn{1}{c|}{\textbf{102}} & 513   & \multicolumn{1}{c|}{}     & \multicolumn{1}{c|}{619}           & \multicolumn{1}{c|}{\textbf{119}} & 630   & \multicolumn{1}{c|}{}     & \multicolumn{1}{c|}{28.1}          & \multicolumn{1}{c|}{\textbf{46}}  & 376   & \multicolumn{1}{c|}{}     & \multicolumn{1}{c|}{110}           & \multicolumn{1}{c|}{27}          & 453          & \multicolumn{1}{c|}{}          & \multicolumn{1}{c|}{74.4}          & \multicolumn{1}{c|}{\textbf{77}} & 641   \\ \cline{3-23} 
\multicolumn{1}{|c|}{}                           & \multicolumn{1}{c|}{}                           & HPEZ-QPET    & \multicolumn{1}{c|}{}     & \multicolumn{1}{c|}{\textbf{954}}  & \multicolumn{1}{c|}{75}           & 398   & \multicolumn{1}{c|}{}     & \multicolumn{1}{c|}{\textbf{1466}} & \multicolumn{1}{c|}{92}           & 413   & \multicolumn{1}{c|}{}     & \multicolumn{1}{c|}{27.9}          & \multicolumn{1}{c|}{44}           & 336   & \multicolumn{1}{c|}{}     & \multicolumn{1}{c|}{\textbf{154}}  & \multicolumn{1}{c|}{\textbf{26}} & 364          & \multicolumn{1}{c|}{}          & \multicolumn{1}{c|}{\textbf{196}}  & \multicolumn{1}{c|}{71}          & 503   \\ \cline{3-23} 
\multicolumn{1}{|c|}{}                           & \multicolumn{1}{c|}{}                           & SPERR-QPET   & \multicolumn{1}{c|}{}     & \multicolumn{1}{c|}{849}           & \multicolumn{1}{c|}{49}           & 135   & \multicolumn{1}{c|}{}     & \multicolumn{1}{c|}{728}           & \multicolumn{1}{c|}{63}           & 260   & \multicolumn{1}{c|}{}     & \multicolumn{1}{c|}{\textbf{35.6}} & \multicolumn{1}{c|}{22}           & 67    & \multicolumn{1}{c|}{}     & \multicolumn{1}{c|}{134}           & \multicolumn{1}{c|}{21}          & 120          & \multicolumn{1}{c|}{}          & \multicolumn{1}{c|}{11.9}          & \multicolumn{1}{c|}{22}          & 53    \\ \hline
\multicolumn{1}{|c|}{\multirow{8}{*}{$10^{-2}$}} & \multicolumn{1}{c|}{\multirow{8}{*}{$10^{-3}$}} & SZ3-OptZ-R   & \multicolumn{1}{c|}{9}    & \multicolumn{1}{c|}{76.5}          & \multicolumn{1}{c|}{19}           & 505   & \multicolumn{1}{c|}{6}    & \multicolumn{1}{c|}{28}            & \multicolumn{1}{c|}{25}           & 444   & \multicolumn{1}{c|}{3}    & \multicolumn{1}{c|}{5.9}           & \multicolumn{1}{c|}{22}           & 152   & \multicolumn{1}{c|}{12}   & \multicolumn{1}{c|}{32.9}          & \multicolumn{1}{c|}{7}           & 320          & \multicolumn{1}{c|}{12}        & \multicolumn{1}{c|}{5.4}           & \multicolumn{1}{c|}{7}           & 144   \\ \cline{3-23} 
\multicolumn{1}{|c|}{}                           & \multicolumn{1}{c|}{}                           & HPEZ-OptZ-R  & \multicolumn{1}{c|}{9}    & \multicolumn{1}{c|}{98.8}          & \multicolumn{1}{c|}{14}           & 381   & \multicolumn{1}{c|}{6}    & \multicolumn{1}{c|}{30.6}          & \multicolumn{1}{c|}{18}           & 322   & \multicolumn{1}{c|}{3}    & \multicolumn{1}{c|}{6.2}           & \multicolumn{1}{c|}{21}           & 183   & \multicolumn{1}{c|}{12}   & \multicolumn{1}{c|}{43.9}          & \multicolumn{1}{c|}{6}           & 320          & \multicolumn{1}{c|}{12}        & \multicolumn{1}{c|}{4.7}           & \multicolumn{1}{c|}{7}           & 154   \\ \cline{3-23} 
\multicolumn{1}{|c|}{}                           & \multicolumn{1}{c|}{}                           & SPERR-OptZ-R & \multicolumn{1}{c|}{10}   & \multicolumn{1}{c|}{108}           & \multicolumn{1}{c|}{5}            & 126   & \multicolumn{1}{c|}{10}   & \multicolumn{1}{c|}{39.6}          & \multicolumn{1}{c|}{7}            & 136   & \multicolumn{1}{c|}{15}   & \multicolumn{1}{c|}{6.1}           & \multicolumn{1}{c|}{2}            & 38    & \multicolumn{1}{c|}{12}   & \multicolumn{1}{c|}{52}            & \multicolumn{1}{c|}{3}           & 99           & \multicolumn{1}{c|}{12}        & \multicolumn{1}{c|}{5.8}           & \multicolumn{1}{c|}{3}           & 39    \\ \cline{3-23} 
\multicolumn{1}{|c|}{}                           & \multicolumn{1}{c|}{}                           & QoI-SZ3      & \multicolumn{1}{c|}{}     & \multicolumn{1}{c|}{169}           & \multicolumn{1}{c|}{35}           & 86    & \multicolumn{1}{c|}{}     & \multicolumn{1}{c|}{N/A}           & \multicolumn{1}{c|}{N/A}          & N/A   & \multicolumn{1}{c|}{}     & \multicolumn{1}{c|}{7.4}           & \multicolumn{1}{c|}{24}           & 112   & \multicolumn{1}{c|}{}     & \multicolumn{1}{c|}{N/A}           & \multicolumn{1}{c|}{N/A}         & N/A          & \multicolumn{1}{c|}{}          & \multicolumn{1}{c|}{N/A}           & \multicolumn{1}{c|}{N/A}         & N/A   \\ \cline{3-23} 
\multicolumn{1}{|c|}{}                           & \multicolumn{1}{c|}{}                           & QoI-HPEZ     & \multicolumn{1}{c|}{}     & \multicolumn{1}{c|}{206}           & \multicolumn{1}{c|}{30}           & 70    & \multicolumn{1}{c|}{}     & \multicolumn{1}{c|}{N/A}           & \multicolumn{1}{c|}{N/A}          & N/A   & \multicolumn{1}{c|}{}     & \multicolumn{1}{c|}{8}             & \multicolumn{1}{c|}{20}           & 123   & \multicolumn{1}{c|}{}     & \multicolumn{1}{c|}{N/A}           & \multicolumn{1}{c|}{N/A}         & N/A          & \multicolumn{1}{c|}{}          & \multicolumn{1}{c|}{N/A}           & \multicolumn{1}{c|}{N/A}         & N/A   \\ \cline{3-23} 
\multicolumn{1}{|c|}{}                           & \multicolumn{1}{c|}{}                           & SZ3-QPET     & \multicolumn{1}{c|}{}     & \multicolumn{1}{c|}{181}           & \multicolumn{1}{c|}{\textbf{76}}  & 332   & \multicolumn{1}{c|}{}     & \multicolumn{1}{c|}{244}           & \multicolumn{1}{c|}{\textbf{88}}  & 505   & \multicolumn{1}{c|}{}     & \multicolumn{1}{c|}{8.8}           & \multicolumn{1}{c|}{33}           & 126   & \multicolumn{1}{c|}{}     & \multicolumn{1}{c|}{34.4}          & \multicolumn{1}{c|}{\textbf{25}} & 288          & \multicolumn{1}{c|}{}          & \multicolumn{1}{c|}{18.1}          & \multicolumn{1}{c|}{\textbf{57}} & 350   \\ \cline{3-23} 
\multicolumn{1}{|c|}{}                           & \multicolumn{1}{c|}{}                           & HPEZ-QPET    & \multicolumn{1}{c|}{}     & \multicolumn{1}{c|}{196}           & \multicolumn{1}{c|}{64}           & 276   & \multicolumn{1}{c|}{}     & \multicolumn{1}{c|}{\textbf{320}}  & \multicolumn{1}{c|}{78}           & 405   & \multicolumn{1}{c|}{}     & \multicolumn{1}{c|}{\textbf{9.2}}  & \multicolumn{1}{c|}{\textbf{33}}  & 128   & \multicolumn{1}{c|}{}     & \multicolumn{1}{c|}{44.5}          & \multicolumn{1}{c|}{24}          & 253          & \multicolumn{1}{c|}{}          & \multicolumn{1}{c|}{\textbf{23.4}} & \multicolumn{1}{c|}{52}          & 297   \\ \cline{3-23} 
\multicolumn{1}{|c|}{}                           & \multicolumn{1}{c|}{}                           & SPERR-QPET   & \multicolumn{1}{c|}{}     & \multicolumn{1}{c|}{\textbf{224}}  & \multicolumn{1}{c|}{47}           & 128   & \multicolumn{1}{c|}{}     & \multicolumn{1}{c|}{210}           & \multicolumn{1}{c|}{59}           & 219   & \multicolumn{1}{c|}{}     & \multicolumn{1}{c|}{9.1}           & \multicolumn{1}{c|}{18}           & 40    & \multicolumn{1}{c|}{}     & \multicolumn{1}{c|}{\textbf{52.5}} & \multicolumn{1}{c|}{20}          & 100          & \multicolumn{1}{c|}{}          & \multicolumn{1}{c|}{5.9}           & \multicolumn{1}{c|}{17}          & 35    \\ \hline
\multicolumn{3}{|c|}{QoI}                                                                                         & \multicolumn{4}{c|}{$Q(x)=x^2$}                                                                            & \multicolumn{4}{c|}{$Q(x)=x^3$}                                                                            & \multicolumn{4}{c|}{$Q(x)=\log_2x$}                                                                        & \multicolumn{4}{c|}{$Q(x)=\sin{10x}$}                                                                            & \multicolumn{4}{c|}{$Q(x)=\tanh{x}$}                                                                           \\ \hline
\multicolumn{3}{|c|}{Data field}                                                                                  & \multicolumn{4}{c|}{Miranda-Density}                                                                       & \multicolumn{4}{c|}{Hurricane-Cloud}                                                                       & \multicolumn{4}{c|}{Scale-LetKF-T}                                                                         & \multicolumn{4}{c|}{Miranda-Viscocity}                                                                           & \multicolumn{4}{c|}{NYX-Velocity\_x}                                                                           \\ \hline
\multicolumn{1}{|c|}{$\epsilon$}                 & \multicolumn{1}{c|}{$\tau$}                     & Compressor   & \multicolumn{1}{c|}{\#It} & \multicolumn{1}{c|}{CR}            & \multicolumn{1}{c|}{$T_c$}        & $T_d$ & \multicolumn{1}{c|}{\#It} & \multicolumn{1}{c|}{CR}            & \multicolumn{1}{c|}{$T_c$}        & $T_d$ & \multicolumn{1}{c|}{\#It} & \multicolumn{1}{c|}{CR}            & \multicolumn{1}{c|}{$T_c$}        & $T_d$ & \multicolumn{1}{c|}{\#It} & \multicolumn{1}{c|}{CR}            & \multicolumn{1}{c|}{$T_c$}       & $T_d$        & \multicolumn{1}{c|}{\#It}      & \multicolumn{1}{c|}{CR}            & \multicolumn{1}{c|}{$T_c$}       & $T_d$ \\ \hline
\multicolumn{1}{|c|}{\multirow{8}{*}{$10^{-3}$}} & \multicolumn{1}{c|}{\multirow{8}{*}{$10^{-3}$}} & SZ3-OptZ-R   & \multicolumn{1}{c|}{4}    & \multicolumn{1}{c|}{87.3}          & \multicolumn{1}{c|}{37}           & 379   & \multicolumn{1}{c|}{8}    & \multicolumn{1}{c|}{49.3}          & \multicolumn{1}{c|}{23}           & 596   & \multicolumn{1}{c|}{3}    & \multicolumn{1}{c|}{73.9}          & \multicolumn{1}{c|}{27}           & 727   & \multicolumn{1}{c|}{8}    & \multicolumn{1}{c|}{47.4}          & \multicolumn{1}{c|}{10}          & 343          & \multicolumn{1}{c|}{17}        & \multicolumn{1}{c|}{1.2}           & \multicolumn{1}{c|}{5}           & 223   \\ \cline{3-23} 
\multicolumn{1}{|c|}{}                           & \multicolumn{1}{c|}{}                           & HPEZ-OptZ-R  & \multicolumn{1}{c|}{4}    & \multicolumn{1}{c|}{112.7}         & \multicolumn{1}{c|}{33}           & 407   & \multicolumn{1}{c|}{4}    & \multicolumn{1}{c|}{53}            & \multicolumn{1}{c|}{40}           & 502   & \multicolumn{1}{c|}{3}    & \multicolumn{1}{c|}{84.3}          & \multicolumn{1}{c|}{21}           & 460   & \multicolumn{1}{c|}{8}    & \multicolumn{1}{c|}{58.9}          & \multicolumn{1}{c|}{9}           & 411          & \multicolumn{1}{c|}{8}         & \multicolumn{1}{c|}{1.2}           & \multicolumn{1}{c|}{11}          & 235   \\ \cline{3-23} 
\multicolumn{1}{|c|}{}                           & \multicolumn{1}{c|}{}                           & SPERR-OptZ-R & \multicolumn{1}{c|}{4}    & \multicolumn{1}{c|}{115}           & \multicolumn{1}{c|}{12}           & 116   & \multicolumn{1}{c|}{5}    & \multicolumn{1}{c|}{26.4}          & \multicolumn{1}{c|}{10}           & 98    & \multicolumn{1}{c|}{3}    & \multicolumn{1}{c|}{142}           & \multicolumn{1}{c|}{15}           & 154   & \multicolumn{1}{c|}{8}    & \multicolumn{1}{c|}{63.4}          & \multicolumn{1}{c|}{5}           & 104          & \multicolumn{1}{c|}{7}         & \multicolumn{1}{c|}{1.3}           & \multicolumn{1}{c|}{3}           & 15    \\ \cline{3-23} 
\multicolumn{1}{|c|}{}                           & \multicolumn{1}{c|}{}                           & QoI-SZ3      & \multicolumn{1}{c|}{}     & \multicolumn{1}{c|}{80.2}          & \multicolumn{1}{c|}{39}           & 84    & \multicolumn{1}{c|}{}     & \multicolumn{1}{c|}{N/A}           & \multicolumn{1}{c|}{N/A}          & N/A   & \multicolumn{1}{c|}{}     & \multicolumn{1}{c|}{67.3}          & \multicolumn{1}{c|}{30}           & 85    & \multicolumn{1}{c|}{}     & \multicolumn{1}{c|}{N/A}           & \multicolumn{1}{c|}{N/A}         & N/A          & \multicolumn{1}{c|}{}          & \multicolumn{1}{c|}{N/A}           & \multicolumn{1}{c|}{N/A}         & N/A   \\ \cline{3-23} 
\multicolumn{1}{|c|}{}                           & \multicolumn{1}{c|}{}                           & QoI-HPEZ     & \multicolumn{1}{c|}{}     & \multicolumn{1}{c|}{90.9}          & \multicolumn{1}{c|}{36}           & 73    & \multicolumn{1}{c|}{}     & \multicolumn{1}{c|}{N/A}           & \multicolumn{1}{c|}{N/A}          & N/A   & \multicolumn{1}{c|}{}     & \multicolumn{1}{c|}{95.8}          & \multicolumn{1}{c|}{28}           & 77    & \multicolumn{1}{c|}{}     & \multicolumn{1}{c|}{N/A}           & \multicolumn{1}{c|}{N/A}         & N/A          & \multicolumn{1}{c|}{}          & \multicolumn{1}{c|}{N/A}           & \multicolumn{1}{c|}{N/A}         & N/A   \\ \cline{3-23} 
\multicolumn{1}{|c|}{}                           & \multicolumn{1}{c|}{}                           & SZ3-QPET     & \multicolumn{1}{c|}{}     & \multicolumn{1}{c|}{96.4}          & \multicolumn{1}{c|}{\textbf{120}} & 450   & \multicolumn{1}{c|}{}     & \multicolumn{1}{c|}{\textbf{69.3}} & \multicolumn{1}{c|}{\textbf{97}}  & 353   & \multicolumn{1}{c|}{}     & \multicolumn{1}{c|}{88.6}          & \multicolumn{1}{c|}{\textbf{106}} & 391   & \multicolumn{1}{c|}{}     & \multicolumn{1}{c|}{51.2}          & \multicolumn{1}{c|}{\textbf{31}} & 400          & \multicolumn{1}{c|}{}          & \multicolumn{1}{c|}{86.5}          & \multicolumn{1}{c|}{\textbf{72}} & 420   \\ \cline{3-23} 
\multicolumn{1}{|c|}{}                           & \multicolumn{1}{c|}{}                           & HPEZ-QPET    & \multicolumn{1}{c|}{}     & \multicolumn{1}{c|}{\textbf{121}}  & \multicolumn{1}{c|}{108}          & 390   & \multicolumn{1}{c|}{}     & \multicolumn{1}{c|}{62.1}          & \multicolumn{1}{c|}{82}           & 329   & \multicolumn{1}{c|}{}     & \multicolumn{1}{c|}{105}           & \multicolumn{1}{c|}{46}           & 329   & \multicolumn{1}{c|}{}     & \multicolumn{1}{c|}{63.4}          & \multicolumn{1}{c|}{29}          & 327          & \multicolumn{1}{c|}{}          & \multicolumn{1}{c|}{\textbf{89.2}} & \multicolumn{1}{c|}{56}          & 291   \\ \cline{3-23} 
\multicolumn{1}{|c|}{}                           & \multicolumn{1}{c|}{}                           & SPERR-QPET   & \multicolumn{1}{c|}{}     & \multicolumn{1}{c|}{118}           & \multicolumn{1}{c|}{44}           & 116   & \multicolumn{1}{c|}{}     & \multicolumn{1}{c|}{36.3}          & \multicolumn{1}{c|}{44}           & 114   & \multicolumn{1}{c|}{}     & \multicolumn{1}{c|}{\textbf{158}}  & \multicolumn{1}{c|}{33}           & 131   & \multicolumn{1}{c|}{}     & \multicolumn{1}{c|}{\textbf{65.2}} & \multicolumn{1}{c|}{22}          & 106          & \multicolumn{1}{c|}{}          & \multicolumn{1}{c|}{72}            & \multicolumn{1}{c|}{32}          & 81    \\ \hline
\multicolumn{1}{|c|}{\multirow{8}{*}{$10^{-4}$}} & \multicolumn{1}{c|}{\multirow{8}{*}{$10^{-4}$}} & SZ3-OptZ-R   & \multicolumn{1}{c|}{4}    & \multicolumn{1}{c|}{31.1}          & \multicolumn{1}{c|}{35}           & 317   & \multicolumn{1}{c|}{5}    & \multicolumn{1}{c|}{29.5}          & \multicolumn{1}{c|}{35}           & 502   & \multicolumn{1}{c|}{3}    & \multicolumn{1}{c|}{16}            & \multicolumn{1}{c|}{21}           & 202   & \multicolumn{1}{c|}{8}    & \multicolumn{1}{c|}{21.5}          & \multicolumn{1}{c|}{10}          & 288          & \multicolumn{1}{c|}{3}         & \multicolumn{1}{c|}{1.2}           & \multicolumn{1}{c|}{30}          & 274   \\ \cline{3-23} 
\multicolumn{1}{|c|}{}                           & \multicolumn{1}{c|}{}                           & HPEZ-OptZ-R  & \multicolumn{1}{c|}{4}    & \multicolumn{1}{c|}{36.1}          & \multicolumn{1}{c|}{31}           & 343   & \multicolumn{1}{c|}{4}    & \multicolumn{1}{c|}{28.2}          & \multicolumn{1}{c|}{34}           & 433   & \multicolumn{1}{c|}{3}    & \multicolumn{1}{c|}{15.6}          & \multicolumn{1}{c|}{20}           & 279   & \multicolumn{1}{c|}{8}    & \multicolumn{1}{c|}{25.5}          & \multicolumn{1}{c|}{9}           & 294          & \multicolumn{1}{c|}{3}         & \multicolumn{1}{c|}{1.2}           & \multicolumn{1}{c|}{25}          & 195   \\ \cline{3-23} 
\multicolumn{1}{|c|}{}                           & \multicolumn{1}{c|}{}                           & SPERR-OptZ-R & \multicolumn{1}{c|}{4}    & \multicolumn{1}{c|}{49.2}          & \multicolumn{1}{c|}{11}           & 97    & \multicolumn{1}{c|}{2}    & \multicolumn{1}{c|}{14.3}          & \multicolumn{1}{c|}{21}           & 73    & \multicolumn{1}{c|}{3}    & \multicolumn{1}{c|}{28.1}          & \multicolumn{1}{c|}{12}           & 96    & \multicolumn{1}{c|}{8}    & \multicolumn{1}{c|}{31.5}          & \multicolumn{1}{c|}{4}           & 86           & \multicolumn{1}{c|}{3}         & \multicolumn{1}{c|}{1.2}           & \multicolumn{1}{c|}{6}           & 14    \\ \cline{3-23} 
\multicolumn{1}{|c|}{}                           & \multicolumn{1}{c|}{}                           & QoI-SZ3      & \multicolumn{1}{c|}{}     & \multicolumn{1}{c|}{29.5}          & \multicolumn{1}{c|}{39}           & 80    & \multicolumn{1}{c|}{}     & \multicolumn{1}{c|}{N/A}           & \multicolumn{1}{c|}{N/A}          & N/A   & \multicolumn{1}{c|}{}     & \multicolumn{1}{c|}{13.2}          & \multicolumn{1}{c|}{29}           & 75    & \multicolumn{1}{c|}{}     & \multicolumn{1}{c|}{N/A}           & \multicolumn{1}{c|}{N/A}         & N/A          & \multicolumn{1}{c|}{}          & \multicolumn{1}{c|}{N/A}           & \multicolumn{1}{c|}{N/A}         & N/A   \\ \cline{3-23} 
\multicolumn{1}{|c|}{}                           & \multicolumn{1}{c|}{}                           & QoI-HPEZ     & \multicolumn{1}{c|}{}     & \multicolumn{1}{c|}{33.4}          & \multicolumn{1}{c|}{35}           & 71    & \multicolumn{1}{c|}{}     & \multicolumn{1}{c|}{N/A}           & \multicolumn{1}{c|}{N/A}          & N/A   & \multicolumn{1}{c|}{}     & \multicolumn{1}{c|}{15.1}          & \multicolumn{1}{c|}{28}           & 70    & \multicolumn{1}{c|}{}     & \multicolumn{1}{c|}{N/A}           & \multicolumn{1}{c|}{N/A}         & N/A          & \multicolumn{1}{c|}{}          & \multicolumn{1}{c|}{N/A}           & \multicolumn{1}{c|}{N/A}         & N/A   \\ \cline{3-23} 
\multicolumn{1}{|c|}{}                           & \multicolumn{1}{c|}{}                           & SZ3-QPET     & \multicolumn{1}{c|}{}     & \multicolumn{1}{c|}{32.4}          & \multicolumn{1}{c|}{\textbf{116}} & 369   & \multicolumn{1}{c|}{}     & \multicolumn{1}{c|}{\textbf{36.6}} & \multicolumn{1}{c|}{\textbf{98}}  & 329   & \multicolumn{1}{c|}{}     & \multicolumn{1}{c|}{16.6}          & \multicolumn{1}{c|}{\textbf{45}}  & 192   & \multicolumn{1}{c|}{}     & \multicolumn{1}{c|}{22.2}          & \multicolumn{1}{c|}{\textbf{28}} & 267          & \multicolumn{1}{c|}{}          & \multicolumn{1}{c|}{15.3}          & \multicolumn{1}{c|}{\textbf{58}} & 202   \\ \cline{3-23} 
\multicolumn{1}{|c|}{}                           & \multicolumn{1}{c|}{}                           & HPEZ-QPET    & \multicolumn{1}{c|}{}     & \multicolumn{1}{c|}{37}            & \multicolumn{1}{c|}{106}          & 327   & \multicolumn{1}{c|}{}     & \multicolumn{1}{c|}{28.2}          & \multicolumn{1}{c|}{70}           & 258   & \multicolumn{1}{c|}{}     & \multicolumn{1}{c|}{16.7}          & \multicolumn{1}{c|}{44}           & 216   & \multicolumn{1}{c|}{}     & \multicolumn{1}{c|}{26.4}          & \multicolumn{1}{c|}{27}          & 229          & \multicolumn{1}{c|}{}          & \multicolumn{1}{c|}{\textbf{16.1}} & \multicolumn{1}{c|}{54}          & 183   \\ \cline{3-23} 
\multicolumn{1}{|c|}{}                           & \multicolumn{1}{c|}{}                           & SPERR-QPET   & \multicolumn{1}{c|}{}     & \multicolumn{1}{c|}{\textbf{50.2}} & \multicolumn{1}{c|}{39}           & 98    & \multicolumn{1}{c|}{}     & \multicolumn{1}{c|}{17.3}          & \multicolumn{1}{c|}{34}           & 79    & \multicolumn{1}{c|}{}     & \multicolumn{1}{c|}{\textbf{30}}   & \multicolumn{1}{c|}{26}           & 85    & \multicolumn{1}{c|}{}     & \multicolumn{1}{c|}{\textbf{32.1}} & \multicolumn{1}{c|}{20}          & 87           & \multicolumn{1}{c|}{}          & \multicolumn{1}{c|}{14.1}          & \multicolumn{1}{c|}{23}          & 51    \\ \hline
\multicolumn{3}{|c|}{QoI}                                                                                         & \multicolumn{4}{c|}{$Q(X)=\frac{1}{n}\sum{x_i}$}                                                           & \multicolumn{4}{c|}{$Q(X)=\frac{1}{n}\sum{x_i^2}$}                                                         & \multicolumn{4}{c|}{$Q(X)=\frac{1}{n}\sum{x_i^3}$}                                                         & \multicolumn{4}{c|}{$Q(x,y,z)=x^2+y^2+z^2$}                                                                      & \multicolumn{4}{c|}{$Q(x,y,z)=\sqrt{x^2+y^2+z^2}$}                                                             \\ \hline
\multicolumn{3}{|c|}{Data field}                                                                                  & \multicolumn{4}{c|}{RTM-3200}                                                                              & \multicolumn{4}{c|}{SegSalt-Pressure3000}                                                                  & \multicolumn{4}{c|}{Scale-LetKF-V}                                                                         & \multicolumn{4}{c|}{Hurricane-UVW}                                                                               & \multicolumn{4}{c|}{Miranda-VXYZ}                                                                              \\ \hline
\multicolumn{1}{|c|}{$\epsilon$}                 & \multicolumn{1}{c|}{$\tau$}                     & Compressor   & \multicolumn{1}{c|}{\#It} & \multicolumn{1}{c|}{CR}            & \multicolumn{1}{c|}{$T_c$}        & $T_d$ & \multicolumn{1}{c|}{\#It} & \multicolumn{1}{c|}{CR}            & \multicolumn{1}{c|}{$T_c$}        & $T_d$ & \multicolumn{1}{c|}{\#It} & \multicolumn{1}{c|}{CR}            & \multicolumn{1}{c|}{$T_c$}        & $T_d$ & \multicolumn{1}{c|}{\#It} & \multicolumn{1}{c|}{CR}            & \multicolumn{1}{c|}{$T_c$}       & $T_d$        & \multicolumn{1}{c|}{\#It}      & \multicolumn{1}{c|}{CR}            & \multicolumn{1}{c|}{$T_c$}       & $T_d$ \\ \hline
\multicolumn{1}{|c|}{\multirow{9}{*}{$10^{-2}$}} & \multicolumn{1}{c|}{\multirow{9}{*}{$10^{-3}$}} & SZ3-OptZ-R   & \multicolumn{1}{c|}{7}    & \multicolumn{1}{c|}{76.1}          & \multicolumn{1}{c|}{20}           & 617   & \multicolumn{1}{c|}{6}    & \multicolumn{1}{c|}{66.7}          & \multicolumn{1}{c|}{24}           & 576   & \multicolumn{1}{c|}{14}   & \multicolumn{1}{c|}{37}            & \multicolumn{1}{c|}{11}           & 605   & \multicolumn{1}{c|}{10}   & \multicolumn{1}{c|}{13}            & \multicolumn{1}{c|}{11}          & 270          & \multicolumn{1}{c|}{8}         & \multicolumn{1}{c|}{122}           & \multicolumn{1}{c|}{14}          & 393   \\ \cline{3-23} 
\multicolumn{1}{|c|}{}                           & \multicolumn{1}{c|}{}                           & HPEZ-OptZ-R  & \multicolumn{1}{c|}{7}    & \multicolumn{1}{c|}{99.2}          & \multicolumn{1}{c|}{14}           & 397   & \multicolumn{1}{c|}{7}    & \multicolumn{1}{c|}{115}           & \multicolumn{1}{c|}{15}           & 388   & \multicolumn{1}{c|}{9}    & \multicolumn{1}{c|}{66.8}          & \multicolumn{1}{c|}{13}           & 434   & \multicolumn{1}{c|}{10}   & \multicolumn{1}{c|}{14}            & \multicolumn{1}{c|}{8}           & 260          & \multicolumn{1}{c|}{8}         & \multicolumn{1}{c|}{179}           & \multicolumn{1}{c|}{13}          & 411   \\ \cline{3-23} 
\multicolumn{1}{|c|}{}                           & \multicolumn{1}{c|}{}                           & SPERR-OptZ-R & \multicolumn{1}{c|}{5}    & \multicolumn{1}{c|}{155.6}         & \multicolumn{1}{c|}{16}           & 205   & \multicolumn{1}{c|}{6}    & \multicolumn{1}{c|}{126}           & \multicolumn{1}{c|}{9}            & 129   & \multicolumn{1}{c|}{7}    & \multicolumn{1}{c|}{94}            & \multicolumn{1}{c|}{9}            & 142   & \multicolumn{1}{c|}{8}    & \multicolumn{1}{c|}{20.4}          & \multicolumn{1}{c|}{7}           & 48           & \multicolumn{1}{c|}{9}         & \multicolumn{1}{c|}{167}           & \multicolumn{1}{c|}{5}           & 121   \\ \cline{3-23} 
\multicolumn{1}{|c|}{}                           & \multicolumn{1}{c|}{}                           & MGARD-QoI    & \multicolumn{1}{c|}{}     & \multicolumn{1}{c|}{15.7}          & \multicolumn{1}{c|}{3}            & 4     & \multicolumn{1}{c|}{}     & \multicolumn{1}{c|}{N/A}           & \multicolumn{1}{c|}{N/A}          & N/A   & \multicolumn{1}{c|}{}     & \multicolumn{1}{c|}{N/A}           & \multicolumn{1}{c|}{N/A}          & N/A   & \multicolumn{1}{c|}{}     & \multicolumn{1}{c|}{N/A}           & \multicolumn{1}{c|}{N/A}         & N/A          & \multicolumn{1}{c|}{}          & \multicolumn{1}{c|}{N/A}           & \multicolumn{1}{c|}{N/A}         & N/A   \\ \cline{3-23} 
\multicolumn{1}{|c|}{}                           & \multicolumn{1}{c|}{}                           & QoI-SZ3      & \multicolumn{1}{c|}{}     & \multicolumn{1}{c|}{53.1}          & \multicolumn{1}{c|}{32}           & 84    & \multicolumn{1}{c|}{}     & \multicolumn{1}{c|}{104}           & \multicolumn{1}{c|}{31}           & 85    & \multicolumn{1}{c|}{}     & \multicolumn{1}{c|}{N/A}           & \multicolumn{1}{c|}{N/A}          & N/A   & \multicolumn{1}{c|}{}     & \multicolumn{1}{c|}{N/A}           & \multicolumn{1}{c|}{N/A}         & N/A          & \multicolumn{1}{c|}{}          & \multicolumn{1}{c|}{N/A}           & \multicolumn{1}{c|}{N/A}         & N/A   \\ \cline{3-23} 
\multicolumn{1}{|c|}{}                           & \multicolumn{1}{c|}{}                           & QoI-HPEZ     & \multicolumn{1}{c|}{}     & \multicolumn{1}{c|}{55.2}          & \multicolumn{1}{c|}{25}           & 68    & \multicolumn{1}{c|}{}     & \multicolumn{1}{c|}{109}           & \multicolumn{1}{c|}{27}           & 70    & \multicolumn{1}{c|}{}     & \multicolumn{1}{c|}{N/A}           & \multicolumn{1}{c|}{N/A}          & N/A   & \multicolumn{1}{c|}{}     & \multicolumn{1}{c|}{N/A}           & \multicolumn{1}{c|}{N/A}         & N/A          & \multicolumn{1}{c|}{}          & \multicolumn{1}{c|}{N/A}           & \multicolumn{1}{c|}{N/A}         & N/A   \\ \cline{3-23} 
\multicolumn{1}{|c|}{}                           & \multicolumn{1}{c|}{}                           & SZ3-QPET     & \multicolumn{1}{c|}{}     & \multicolumn{1}{c|}{93}            & \multicolumn{1}{c|}{\textbf{97}}  & 430   & \multicolumn{1}{c|}{}     & \multicolumn{1}{c|}{156}           & \multicolumn{1}{c|}{\textbf{80}}  & 455   & \multicolumn{1}{c|}{}     & \multicolumn{1}{c|}{329}           & \multicolumn{1}{c|}{\textbf{71}}  & 339   & \multicolumn{1}{c|}{}     & \multicolumn{1}{c|}{26.6}          & \multicolumn{1}{c|}{\textbf{46}} & \textbf{172} & \multicolumn{1}{c|}{\textbf{}} & \multicolumn{1}{c|}{163}           & \multicolumn{1}{c|}{\textbf{29}} & 497   \\ \cline{3-23} 
\multicolumn{1}{|c|}{}                           & \multicolumn{1}{c|}{}                           & HPEZ-QPET    & \multicolumn{1}{c|}{}     & \multicolumn{1}{c|}{140}           & \multicolumn{1}{c|}{75}           & 361   & \multicolumn{1}{c|}{}     & \multicolumn{1}{c|}{234}           & \multicolumn{1}{c|}{70}           & 369   & \multicolumn{1}{c|}{}     & \multicolumn{1}{c|}{300}           & \multicolumn{1}{c|}{59}           & 288   & \multicolumn{1}{c|}{}     & \multicolumn{1}{c|}{26.8}          & \multicolumn{1}{c|}{43}          & 152          & \multicolumn{1}{c|}{}          & \multicolumn{1}{c|}{\textbf{223}}  & \multicolumn{1}{c|}{28}          & 389   \\ \cline{3-23} 
\multicolumn{1}{|c|}{}                           & \multicolumn{1}{c|}{}                           & SPERR-QPET   & \multicolumn{1}{c|}{}     & \multicolumn{1}{c|}{\textbf{238}}  & \multicolumn{1}{c|}{65}           & 213   & \multicolumn{1}{c|}{}     & \multicolumn{1}{c|}{\textbf{254}}  & \multicolumn{1}{c|}{44}           & 131   & \multicolumn{1}{c|}{}     & \multicolumn{1}{c|}{\textbf{351}}  & \multicolumn{1}{c|}{41}           & 148   & \multicolumn{1}{c|}{}     & \multicolumn{1}{c|}{39}            & \multicolumn{1}{c|}{41}          & 57           & \multicolumn{1}{c|}{}          & \multicolumn{1}{c|}{202}           & \multicolumn{1}{c|}{22}          & 62    \\ \hline
\multicolumn{1}{|c|}{\multirow{9}{*}{$10^{-3}$}} & \multicolumn{1}{c|}{\multirow{9}{*}{$10^{-4}$}} & SZ3-OptZ-R   & \multicolumn{1}{c|}{7}    & \multicolumn{1}{c|}{15.2}          & \multicolumn{1}{c|}{18}           & 341   & \multicolumn{1}{c|}{7}    & \multicolumn{1}{c|}{19.4}          & \multicolumn{1}{c|}{19}           & 353   & \multicolumn{1}{c|}{10}   & \multicolumn{1}{c|}{14.3}          & \multicolumn{1}{c|}{10}           & 190   & \multicolumn{1}{c|}{4}    & \multicolumn{1}{c|}{6.5}           & \multicolumn{1}{c|}{17}          & 132          & \multicolumn{1}{c|}{8}         & \multicolumn{1}{c|}{35.8}          & \multicolumn{1}{c|}{13}          & 325   \\ \cline{3-23} 
\multicolumn{1}{|c|}{}                           & \multicolumn{1}{c|}{}                           & HPEZ-OptZ-R  & \multicolumn{1}{c|}{6}    & \multicolumn{1}{c|}{15.2}          & \multicolumn{1}{c|}{16}           & 256   & \multicolumn{1}{c|}{8}    & \multicolumn{1}{c|}{28.1}          & \multicolumn{1}{c|}{13}           & 314   & \multicolumn{1}{c|}{8}    & \multicolumn{1}{c|}{15.5}          & \multicolumn{1}{c|}{13}           & 286   & \multicolumn{1}{c|}{4}    & \multicolumn{1}{c|}{5.9}           & \multicolumn{1}{c|}{17}          & 172          & \multicolumn{1}{c|}{8}         & \multicolumn{1}{c|}{45.5}          & \multicolumn{1}{c|}{12}          & 363   \\ \cline{3-23} 
\multicolumn{1}{|c|}{}                           & \multicolumn{1}{c|}{}                           & SPERR-OptZ-R & \multicolumn{1}{c|}{6}    & \multicolumn{1}{c|}{24.4}          & \multicolumn{1}{c|}{10}           & 106   & \multicolumn{1}{c|}{9}    & \multicolumn{1}{c|}{30.3}          & \multicolumn{1}{c|}{5}            & 91    & \multicolumn{1}{c|}{13}   & \multicolumn{1}{c|}{22.7}          & \multicolumn{1}{c|}{4}            & 87    & \multicolumn{1}{c|}{8}    & \multicolumn{1}{c|}{8.3}           & \multicolumn{1}{c|}{5}           & 42           & \multicolumn{1}{c|}{9}         & \multicolumn{1}{c|}{58.6}          & \multicolumn{1}{c|}{5}           & 102   \\ \cline{3-23} 
\multicolumn{1}{|c|}{}                           & \multicolumn{1}{c|}{}                           & MGARD-QoI    & \multicolumn{1}{c|}{}     & \multicolumn{1}{c|}{5.9}           & \multicolumn{1}{c|}{3}            & 4     & \multicolumn{1}{c|}{}     & \multicolumn{1}{c|}{N/A}           & \multicolumn{1}{c|}{N/A}          & N/A   & \multicolumn{1}{c|}{}     & \multicolumn{1}{c|}{N/A}           & \multicolumn{1}{c|}{N/A}          & N/A   & \multicolumn{1}{c|}{}     & \multicolumn{1}{c|}{N/A}           & \multicolumn{1}{c|}{N/A}         & N/A          & \multicolumn{1}{c|}{}          & \multicolumn{1}{c|}{N/A}           & \multicolumn{1}{c|}{N/A}         & N/A   \\ \cline{3-23} 
\multicolumn{1}{|c|}{}                           & \multicolumn{1}{c|}{}                           & QoI-SZ3      & \multicolumn{1}{c|}{}     & \multicolumn{1}{c|}{13.6}          & \multicolumn{1}{c|}{31}           & 75    & \multicolumn{1}{c|}{}     & \multicolumn{1}{c|}{34.5}          & \multicolumn{1}{c|}{31}           & 82    & \multicolumn{1}{c|}{}     & \multicolumn{1}{c|}{N/A}           & \multicolumn{1}{c|}{N/A}          & N/A   & \multicolumn{1}{c|}{}     & \multicolumn{1}{c|}{N/A}           & \multicolumn{1}{c|}{N/A}         & N/A          & \multicolumn{1}{c|}{}          & \multicolumn{1}{c|}{N/A}           & \multicolumn{1}{c|}{N/A}         & N/A   \\ \cline{3-23} 
\multicolumn{1}{|c|}{}                           & \multicolumn{1}{c|}{}                           & QoI-HPEZ     & \multicolumn{1}{c|}{}     & \multicolumn{1}{c|}{15.1}          & \multicolumn{1}{c|}{28}           & 62    & \multicolumn{1}{c|}{}     & \multicolumn{1}{c|}{50}            & \multicolumn{1}{c|}{27}           & 65    & \multicolumn{1}{c|}{}     & \multicolumn{1}{c|}{N/A}           & \multicolumn{1}{c|}{N/A}          & N/A   & \multicolumn{1}{c|}{}     & \multicolumn{1}{c|}{N/A}           & \multicolumn{1}{c|}{N/A}         & N/A          & \multicolumn{1}{c|}{}          & \multicolumn{1}{c|}{N/A}           & \multicolumn{1}{c|}{N/A}         & N/A   \\ \cline{3-23} 
\multicolumn{1}{|c|}{}                           & \multicolumn{1}{c|}{}                           & SZ3-QPET     & \multicolumn{1}{c|}{}     & \multicolumn{1}{c|}{19}            & \multicolumn{1}{c|}{\textbf{87}}  & 301   & \multicolumn{1}{c|}{}     & \multicolumn{1}{c|}{49.4}          & \multicolumn{1}{c|}{\textbf{59}}  & 265   & \multicolumn{1}{c|}{}     & \multicolumn{1}{c|}{56.5}          & \multicolumn{1}{c|}{\textbf{64}}  & 304   & \multicolumn{1}{c|}{}     & \multicolumn{1}{c|}{8.7}           & \multicolumn{1}{c|}{\textbf{46}} & \textbf{130} & \multicolumn{1}{c|}{\textbf{}} & \multicolumn{1}{c|}{43.6}          & \multicolumn{1}{c|}{\textbf{29}} & 419   \\ \cline{3-23} 
\multicolumn{1}{|c|}{}                           & \multicolumn{1}{c|}{}                           & HPEZ-QPET    & \multicolumn{1}{c|}{}     & \multicolumn{1}{c|}{20.8}          & \multicolumn{1}{c|}{71}           & 251   & \multicolumn{1}{c|}{}     & \multicolumn{1}{c|}{67.2}          & \multicolumn{1}{c|}{54}           & 236   & \multicolumn{1}{c|}{}     & \multicolumn{1}{c|}{82.7}          & \multicolumn{1}{c|}{57}           & 269   & \multicolumn{1}{c|}{}     & \multicolumn{1}{c|}{9.1}           & \multicolumn{1}{c|}{43}          & 124          & \multicolumn{1}{c|}{}          & \multicolumn{1}{c|}{54.3}          & \multicolumn{1}{c|}{28}          & 343   \\ \cline{3-23} 
\multicolumn{1}{|c|}{}                           & \multicolumn{1}{c|}{}                           & SPERR-QPET   & \multicolumn{1}{c|}{}     & \multicolumn{1}{c|}{\textbf{43.4}} & \multicolumn{1}{c|}{50}           & 128   & \multicolumn{1}{c|}{}     & \multicolumn{1}{c|}{\textbf{73.6}} & \multicolumn{1}{c|}{41}           & 116   & \multicolumn{1}{c|}{}     & \multicolumn{1}{c|}{\textbf{83.3}} & \multicolumn{1}{c|}{38}           & 121   & \multicolumn{1}{c|}{}     & \multicolumn{1}{c|}{13}            & \multicolumn{1}{c|}{30}          & 32           & \multicolumn{1}{c|}{}          & \multicolumn{1}{c|}{\textbf{67.4}} & \multicolumn{1}{c|}{21}          & 53    \\ \hline
\end{tabular}
\end{table*}
In Table~\ref{tab:showcase}, on diverse data domains, we present 15 QoI-preserving error-bounded lossy compression showcases for the 10 QoI functions in Table~\ref{tab:qoilist}, leveraging QPET and all available baselines.
We apply different error-bound targets and report both compression ratio and sequential throughputs. 

From those showcases, key findings are: (1) Matching the discussion in Section~\ref{sec:eva:speeds}, QPET-integrated compressors deliver better compression throughputs than all baselines. The QPET-integrated compressor achieves 2x to 10x compression speedups over the parameter-search-based solutions. Moreover, SZ3/HPEZ-QPET has 2x to 3x compression speeds and 3x to 6x decompression speeds over QoI-SZ3/HPEZ. (2) With better throughputs, QPET-integrated compressors bring no worse (and mostly improved) compression ratios over the baselines. On $q(x)=x^2$, under error bounds of $10^{-4}$, QPET gained similar compression ratios with baselines in much shorter time costs. On preserving $q(x)=x^3$ of RTM-3500 and Hurricane-Cloud data (they both feature small absolute values less than 0.002), QPET-integrated compressors achieved 30\% - 700\% compression ratio improvements over the best parameter-search solution. On NYX-Velocity\_x data ranging from zero to tens of millions, general-purpose compressors failed to preserve the QoI $\tanh{x}$ well due to the required point-wise data accuracy extremely diverging among small and large values. In contrast, QPET handled this case excellently. On multivariate QoIs, QPET significantly improved the compression ratio over baselines, as benefited from Theorem~\ref{theo:4}. 
Compared with existing QoI-preserving compressors on basic QoIs, QPET-integrated compressors still achieve noticeable compression ratio gains. On preserving $q(x)=\log_2x$ on Scale-LetKF-T data, SPERR-QPET achieved 100\% CR improvement over QoI-HPEZ under $\epsilon=\tau=10^{-4}$. On preserving $q(X)=\frac{1}{n}\sum x_i^2$ on SegSalt-Pressure-3000 data, SPERR-QPET achieved 133\% CR improvement over QoI-HPEZ under $\tau=10^{-3}$. 
General-purpose compressors usually over-preserve data quality to preserve the QoI error threshold because they can neither apply point-wise data accuracy nor auto-correct outliers. This explains why non-QoI-integrated compressors have limited compression ratios on QoI-preserving compression tasks, even after parameter searches on optimizing the compression ratio. Meanwhile, existing QoI-preserving compressors (like QoI-SZ3/HPEZ and MGARD-QoI) bring significant computational overheads for the QoI-preserving compression tasks. With lightweight error-bound tuning and outlier correction mechanisms, QPET successfully avoided the abovementioned limitations, proving its effectiveness and usability in preserving the QoI.

\subsubsection{Improved Rate-distortions}

\begin{figure}[ht] 
\centering
\hspace{-10mm}
\subfigure[{SegSalt, $Q(x)=x^2$}]
{
\raisebox{-1cm}{\includegraphics[scale=\subfigsize]{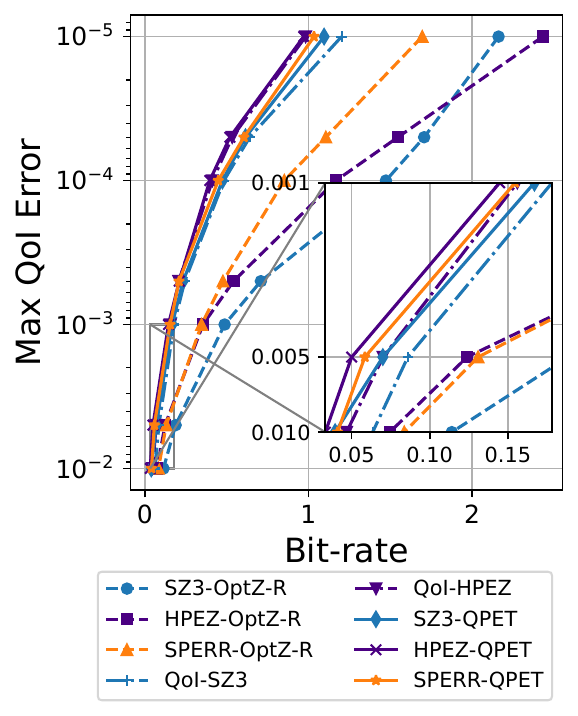}}
}
\hspace{-3mm}
\subfigure[{SCALE-LetKF, $Q(x)=x^2$}]
{
\raisebox{-1cm}{\includegraphics[scale=\subfigsize]{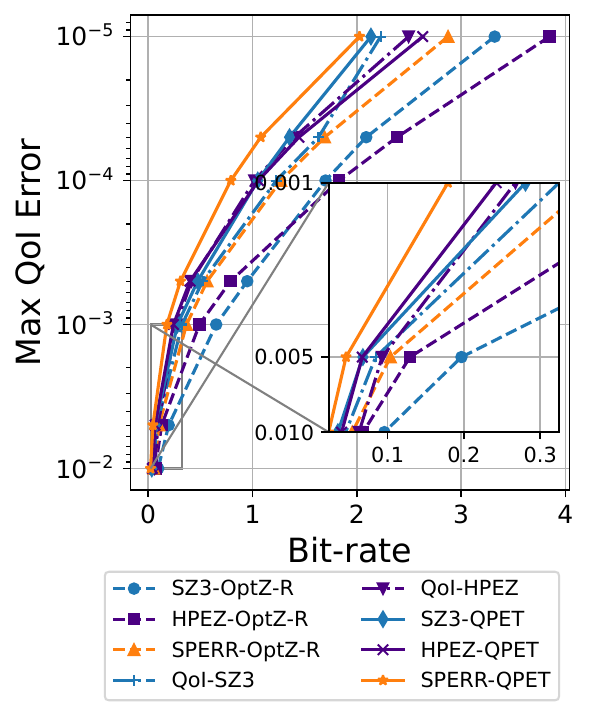}}
}
\hspace{-10mm}

\hspace{-10mm}
\subfigure[{RTM, $Q(x)=x^3$}]
{
\raisebox{-1cm}{\includegraphics[scale=\subfigsize]{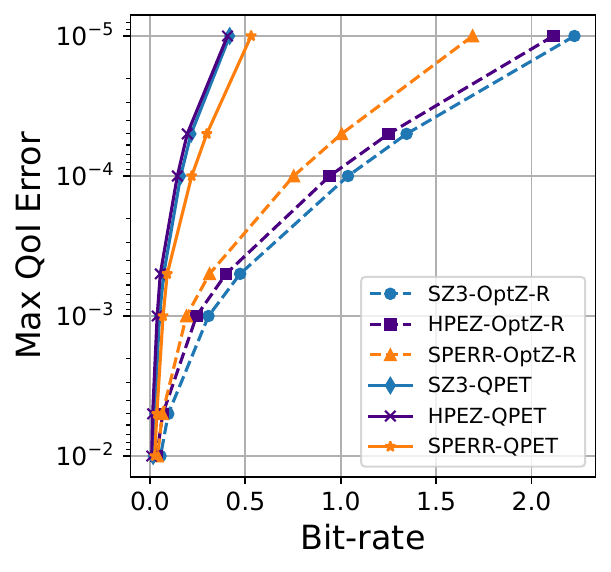}}
}
\hspace{-3mm}
\subfigure[{NYX, $Q(x)=x^3$}]
{
\raisebox{-1cm}{\includegraphics[scale=\subfigsize]{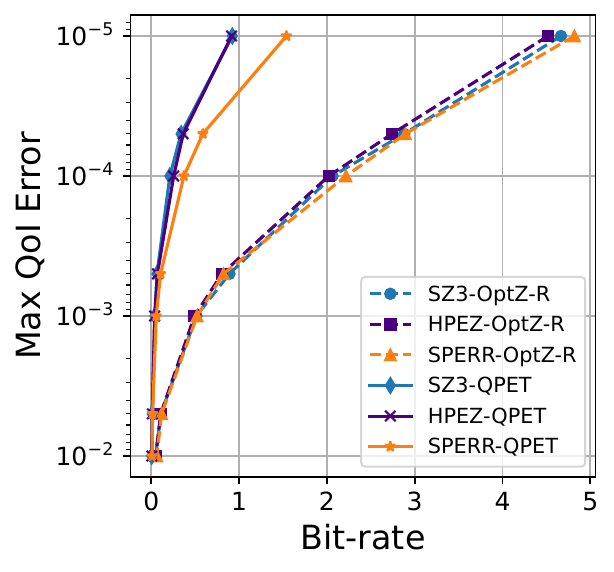}}%
}
\hspace{-10mm}

\hspace{-10mm}
\subfigure[{Miranda, $Q(x)=\sin{10x}$}]
{
\raisebox{-1cm}{\includegraphics[scale=\subfigsize]{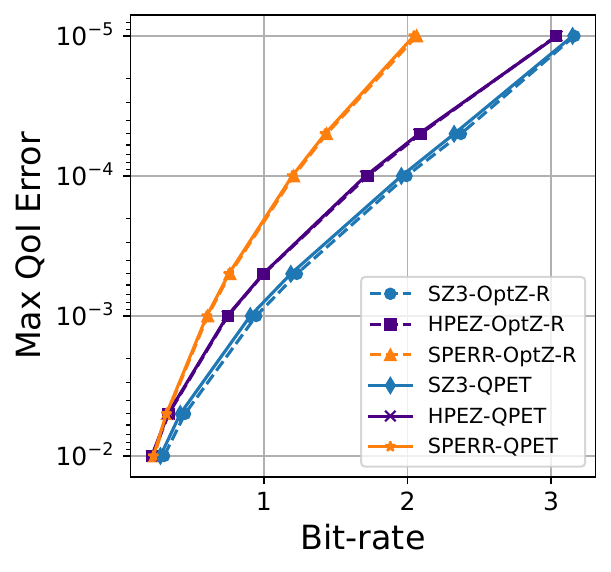}}%
}
\hspace{-3mm}
\subfigure[{Hurricane, $Q(x)=\tanh{x}$}]
{
\raisebox{-1cm}{\includegraphics[scale=\subfigsize]{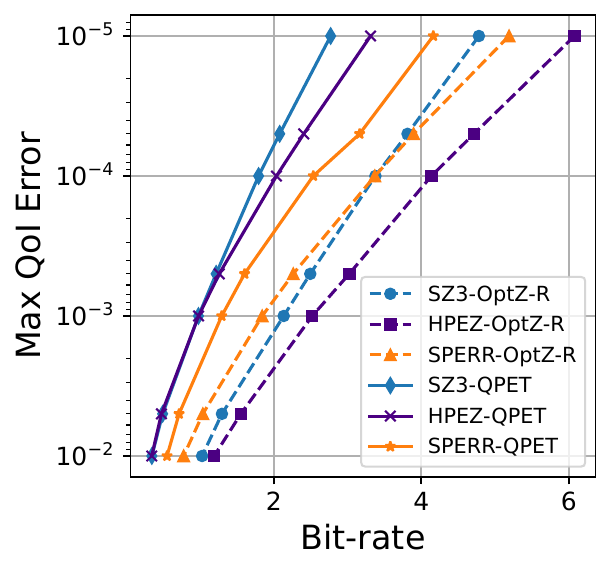}}%
}
\hspace{-10mm}

\caption{Bit rate and max QoI error plots (point-wise QoIs). }
\label{fig:evaluation-bq-pw}
\end{figure}

To comprehensively understand the compression ratio improvement brought about by QPET in QoI-preserving error-bounded data compression, the compressors mentioned above are evaluated on a wide range of QoI functions and error thresholds. Then, we plot the bit rates and maximum QoI errors (relative, absolute value divided by value range), showing that QPET-integrated compressors have substantially improved the compression ratio when achieving the same maximum QoI errors as others. Due to the page limit, we will present the most representative results among the excellent outcomes we have achieved. 

\begin{figure}[ht] 
\centering
\hspace{-10mm}
\subfigure[{Miranda, $Q(X) = \frac{1}{n_b}\sum x$}]
{
\raisebox{-1cm}{\includegraphics[scale=\subfigsize]{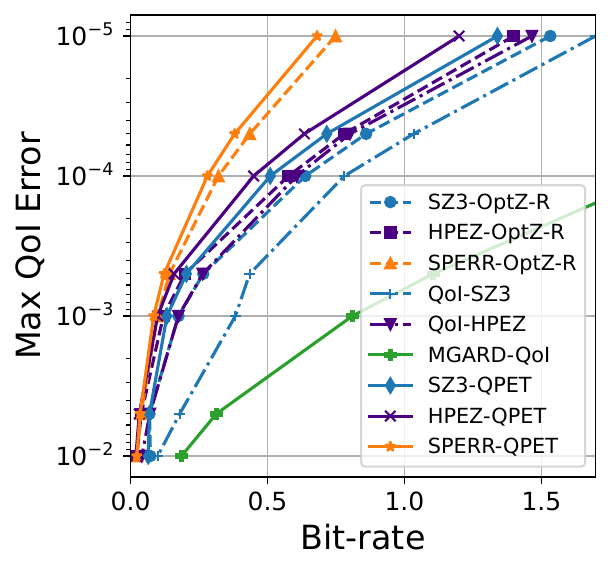}}
}
\hspace{-3mm}
\subfigure[{SegSalt, $Q(X) = \frac{1}{n_b}\sum x$}]
{
\raisebox{-1cm}{\includegraphics[scale=\subfigsize]{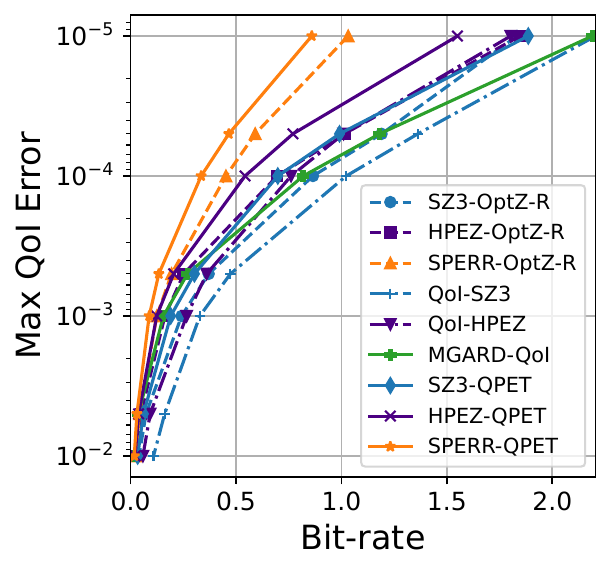}}
}
\hspace{-10mm}

\hspace{-10mm}
\subfigure[{RTM, $Q(X) = \frac{1}{n_b}\sum x^2$}]
{
\raisebox{-1cm}{\includegraphics[scale=\subfigsize]{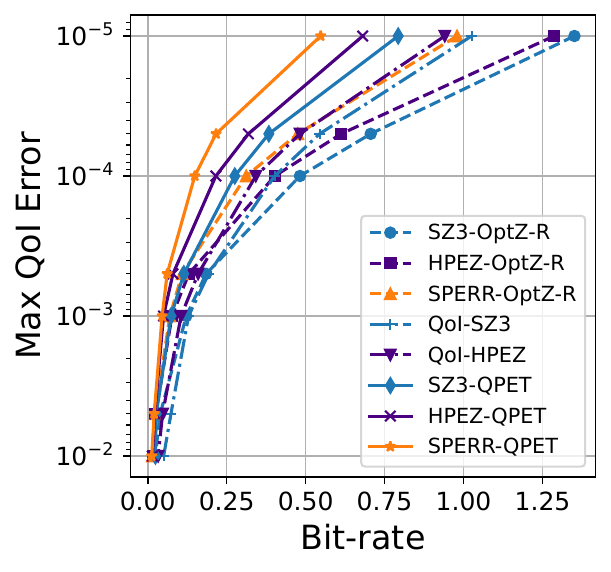}}
}
\hspace{-3mm}
\subfigure[{NYX, $Q(X) = \frac{1}{n_b}\sum x^2$}]
{
\raisebox{-1cm}{\includegraphics[scale=\subfigsize]{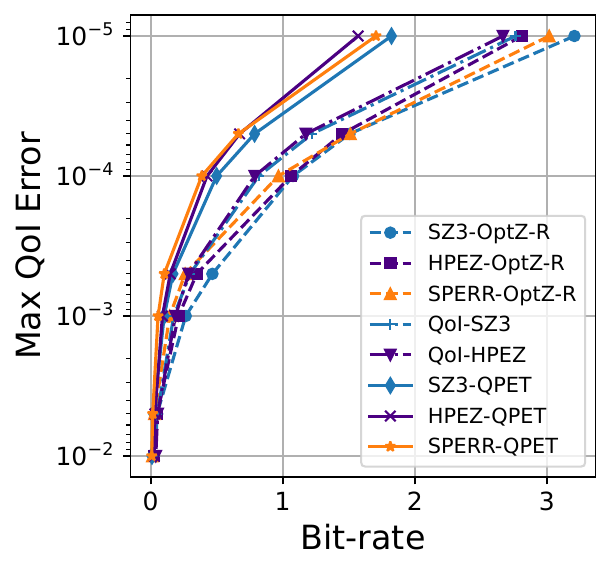}}%
}
\hspace{-10mm}

\hspace{-10mm}
\subfigure[{Hurricane, $Q(X) = \frac{1}{n_b}\sum x^3$}]
{
\raisebox{-1cm}{\includegraphics[scale=\subfigsize]{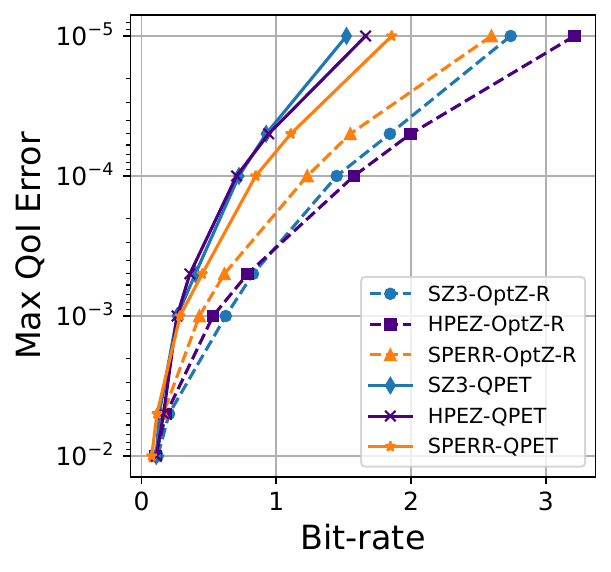}}%
}
\hspace{-3mm}
\subfigure[{Scale-LetKF, $Q(X) = \frac{1}{n_b}\sum x^3$}]
{
\raisebox{-1cm}{\includegraphics[scale=\subfigsize]{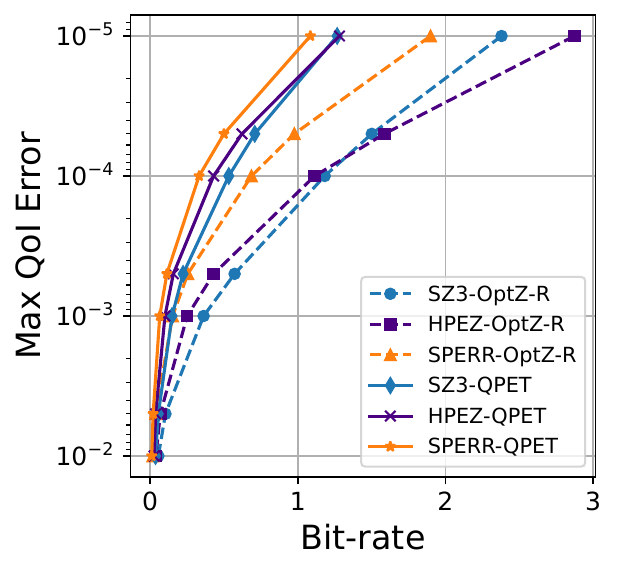}}%
}
\hspace{-10mm}
\caption{Bit rate and max QoI error plots of compressors on Regional QoIs ($n_b = 4^3$, i.e., average on 4x4x4 blocks).}
\label{fig:evaluation-bq-block}
\end{figure}

Regarding different point-wise QoI functions, Figure~\ref{fig:evaluation-bq-pw} presents the bit rate-maximum QoI error plots obtained on 6 data domains. In this figure and all other similar ones in the paper, the results from the same base compressors are plotted as curves of the same color, with QPET-integrated compressors represented by solid lines and baselines by dashed lines. Moreover, the results from QoI-SZ3/HPEZ appear if and only if the corresponding) QoI is supported by them.
In Figure~\ref{fig:evaluation-bq-pw}, QPET-integrated compressors have shown the optimized compression ratio (bit rate) in every test case and achieve substantial improvements in most. For example, on the Scale-LetKF dataset (Figure~\ref{fig:evaluation-bq-pw}~(b)), SPERR-QPET has around 100\%/50\% compression ratio improvements over the best baseline (QoI-SZ3/HPEZ) when the error threshold for the square of data ($Q(x) = x^2$) is set as $1\mathrm{e}{-2}$/$1\mathrm{e}{-3}$ (relative, and all the followings are same). On other QoIs for which existing QoI-preserving data compressors fail to support, such as $x^3$ and $\tanh{x}$ (Figure~\ref{fig:evaluation-bq-pw}~(c, d, f)), QPET makes great use of the point-wise data accuracy to preserve QoIs, significantly outperforming parameter-search-based baselines (*-OptZ-R) by up to over 1000\% compression ratio improvement (on NYX dataset with $Q(x)=x^3$) and higher throughputs. Regarding the preservation of $Q(x)=\sin{10x}$, which is close to a linear function in several local data regions, point-wise data accuracy contributes a limited amount to the compression. Therefore, the compression ratios of QPET-integrated compressors do not outperform parameter-search-based baselines. Nevertheless, QPET-integrated compressors can acquire the same compression ratios at much lower time costs, which is still a critical improvement for high-performance data processing tasks.

\begin{figure}[ht] 
\centering
\hspace{-10mm}
\subfigure[{Miranda, $\sqrt{v_x^2+v_y^2+v_z^2}$}]
{
\raisebox{-1cm}{\includegraphics[scale=\subfigsize]{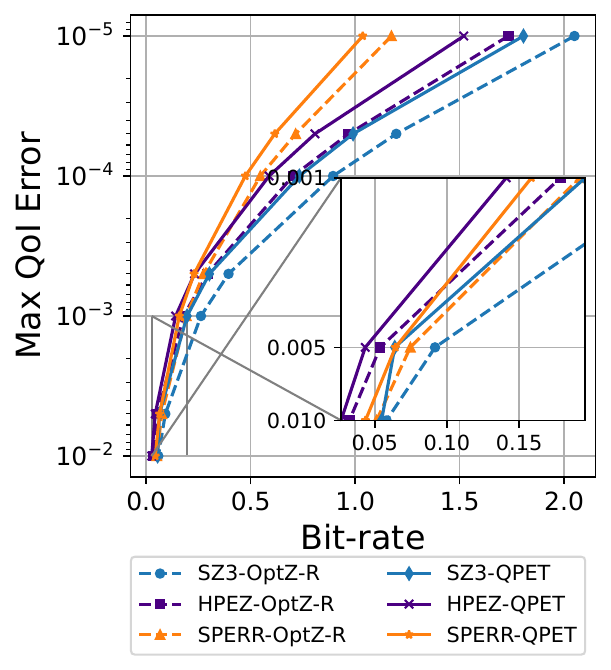}}
}
\hspace{-3mm}
\subfigure[{Hurricane, $\sqrt{u^2+v^2+w^2}$}]
{
\raisebox{-1cm}{\includegraphics[scale=\subfigsize]{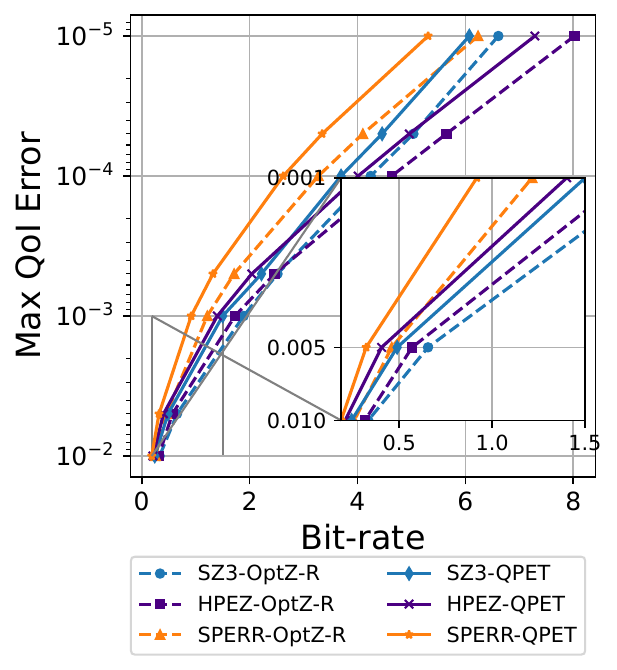}}
}
\hspace{-10mm}

\hspace{-10mm}
\subfigure[{Nyx, $v_x^2+v_y^2+v_z^2$}]
{
\raisebox{-1cm}{\includegraphics[scale=\subfigsize]{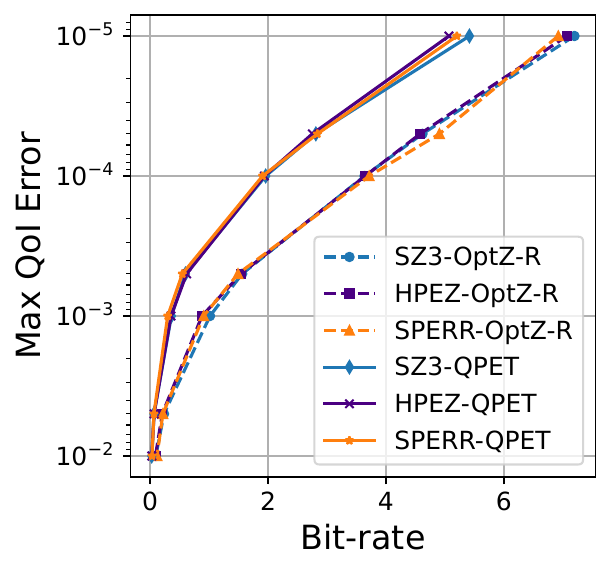}}%
}
\hspace{-3mm}
\subfigure[{SCALE-LetKF, $u^2+v^2+w^2$}]
{
\raisebox{-1cm}{\includegraphics[scale=\subfigsize]{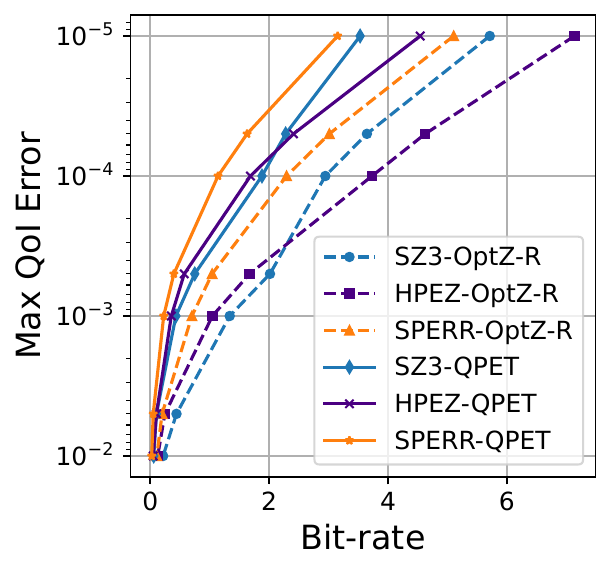}}%
}
\hspace{-10mm}
\caption{Bit rate and max QoI error plots (vector QoIs).}
\label{fig:evaluation-bq-vector}
\end{figure}

Multivariate QoI functions contain two sub-categories: QoI on data blocks within one data field and QoI vector data among multiple fields. In Figure~\ref{fig:evaluation-bq-block}, for better comparison with baselines (particularly QoI-SZ3/HPEZ and MGARD-QoI), we perform evaluations of QPET and the baselines on the average of $x$ (data value itself), $x^2$ (square of the data value), and $x^3$ (cubic of the data value) on partitioned fixed-size data blocks.
In Figure~\ref{fig:evaluation-bq-block}~(a) and (b), we can find that MGARD-QoI and QoI-SZ3 present unstable and suboptimal compression ratios, and QoI-HPEZ cannot outperform HPEZ-QPET and SPERR-QPET with an obvious gap between them. For $Q(X) = \frac{1}{n_b}\sum x^2$ and $Q(X) = \frac{1}{n_b}\sum x^3$, the 3 QPET-integrated compressors are the top-3 of all compressors, outperforming both parameter-search-based baselines and QoI-SZ3/HPEZ. On preserving the $Q(X) = \frac{1}{n_b}\sum x^2$ when compressing the RTM dataset (Figure~\ref{fig:evaluation-bq-block}~(c)), SPERR-QPET gains an aggregated compression ratio of $\approx$ 215 (bit rate of $\approx $0.15) under the QoI error tolerance of $1\mathrm{e}{-4}$, which is 108\% better than the best baseline SPERR-OptZ-R (CR $\approx$ 103). On preserving the $Q(X) = \frac{1}{n_b}\sum x^3$ when compressing the Hurricane dataset (Figure~\ref{fig:evaluation-bq-block}~(e)),  HPEZ-QPET gains aggregated compression ratios of $\approx$ 121/45.4  under the QoI error tolerance of $1\mathrm{e}{-3}$/$1\mathrm{e}{-4}$, which is 200\%/220\% of HPEZ-OptZ-R compression ratios ( $\approx$ 60.9/20.6).
The improvement of compression ratios in those test cases is highly attributed to Theorem~\ref{theo:4} because of the large number of variables (same as the block size 64) in the QoIs so that QPET can apply much higher point-wise QoI ($x$/$x^2$/$x^3$) error thresholds on each data point compared to the overall error threshold.

In Figure~\ref{fig:evaluation-bq-vector}, we also evaluate several vector-style multivariate QoIs. On those vector QoIs, QPET also provides satisfactory compression ratio improvements under all test cases. On the Hurricane dataset, with $Q(x)=\sqrt{u^2+v^2+w^2}$ and $\tau = 1\mathrm{e}{-3}$, SPERR-QPET improves the compression ratio of SPERR-OptZ-R by 25\%. On the NYX dataset, with $Q(x)=u^2+v^2+w^2$ and $\tau$ = $1\mathrm{e}{-3}$/$1\mathrm{e}{-4}$, HPEZ-QPET improves the CR of the baselines by over 140\%/80\%. 

\subsubsection{Ablation study}

\begin{figure}[ht] 
\centering
\hspace{-10mm}
\subfigure[{NYX, $Q(x)=x^3$}]
{
\raisebox{-1cm}{\includegraphics[scale=\subfigsize]{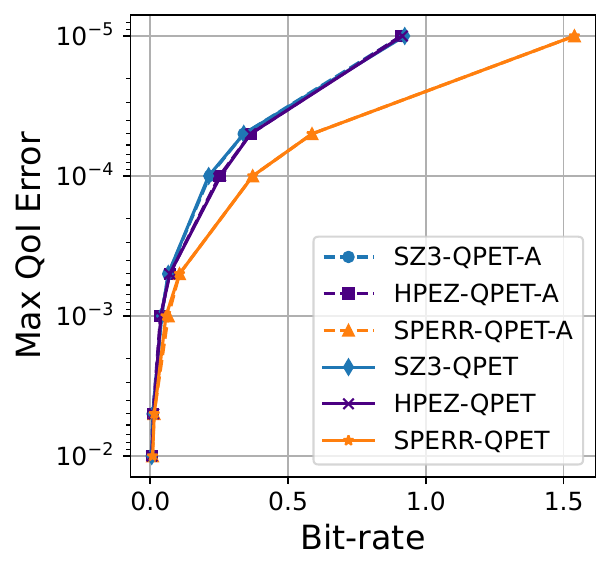}}%
}
\hspace{-3mm}
\subfigure[{Hurricane, $Q(x)=x^3$}]
{
\raisebox{-1cm}{\includegraphics[scale=\subfigsize]{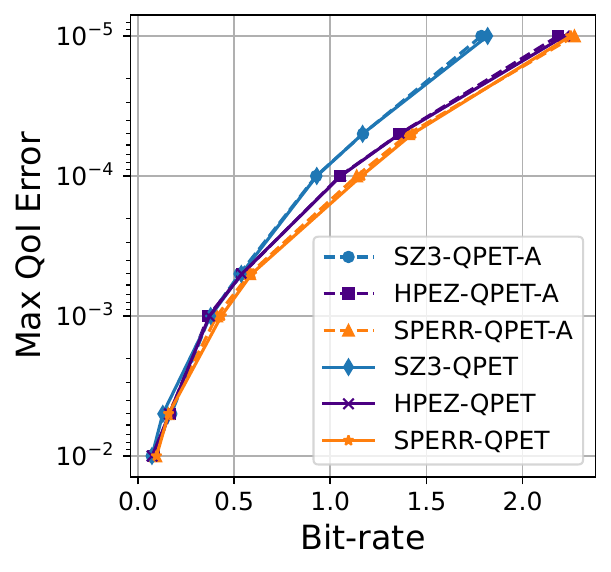}}
}
\hspace{-10mm}

\hspace{-10mm}
\subfigure[{SCALE-LetKF,  $Q(X)=\frac{1}{n_b}\sum x^2$}]
{
\raisebox{-1cm}{\includegraphics[scale=\subfigsize]{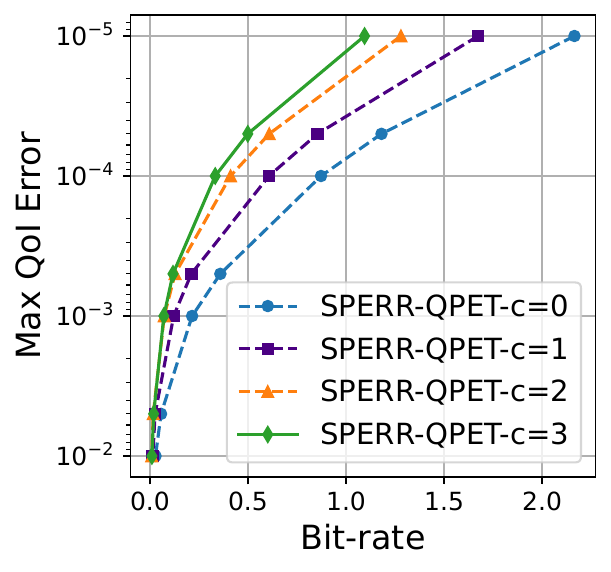}}%
}
\hspace{-3mm}
\subfigure[{Miranda, $Q(x)=x^2$}]
{
\raisebox{-1cm}{\includegraphics[scale=\subfigsize]{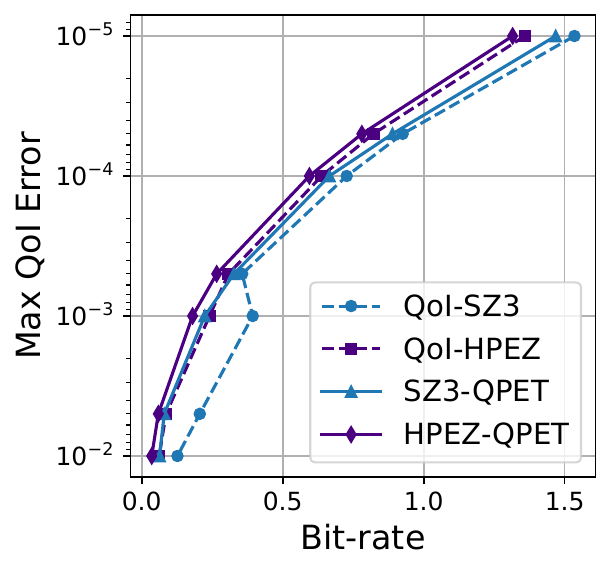}}
}
\hspace{-10mm}
\caption{Ablation study on QPET Components.}
\label{fig:evaluation-bq-abla}
\end{figure}
At the end of our evaluation, we conduct an ablation study of QPET design components, separately validating how different strategies described in Algorithm~\ref{alg:point-wise}, ~\ref{alg:regional}, and ~\ref{alg:globaltune} contribute to the performance of QPET.

On bounding the error of $Q(x)=x^3$, Figure~\ref{fig:evaluation-bq-abla} (a) and (b) compare QPET to its derivation QPET-A, which uses the analytical solution of the inequality $\left|(\epsilon_i+x_i)^3-x_i^3\right| \leq \tau$ instead of Algorithm~\ref{alg:point-wise} to determine the point-wise error bound $\epsilon_0$ for $x_0$. On both datasets, QPET achieves the same compression ratio as QPET-A, indicating that Algorithm~\ref{alg:point-wise} performs well in replacing analytic solutions. This fact indicates that the mechanism of QPET has worked well in determining the compression error bounds.

Next, we investigate the impact of Throrem~\ref{theo:4} and the parameter $c$ by evaluating SPERR-QPET on the SCALE-LetKF with the QoI of the average of $x^2$ on 4$\times$4$\times$4 data blocks with different configurations. Under $c=0$, SPERR-QPET deactivates Theorem~\ref{theo:4}, and larger $c$ brings large error bound estimations from Throrem~\ref{theo:4}. Figure~\ref{fig:evaluation-bq-abla} (c) illustrates the corresponding results. Higher $c$ does improve compression by proposing higher and still trustworthy estimations of point-wise error bounds. As the improvement in compression ratios becomes minor when $c$ reaches 3.0, we selected this value as the default in SPERR-QPET.  

Last, in Figure~\ref{fig:evaluation-bq-abla} (d), we verify the stability and quality of QPET's global error-bound tuning mechanism (Algorithm~\ref{alg:globaltune}). To this end, with $Q(x) = x^2$, we compared SZ3/HPEZ-QPET to QoI-SZ3/HPEZ. On the Miranda dataset, SZ3-QPET resolves the instability of QoI-SZ3 (which presents performance degradation on large error bounds due to suboptimal error bound selection), and HPEZ-QPET provides better compression ratios (5\% $\sim$ 20\% across different QoI error thresholds) than QoI-HPEZ. According to these results and more similar ones we acquired, we can state that Algorithm~\ref{alg:globaltune} achieves state-of-the-art for global error-bound tuning.
\section{Conclusion}
\label{sec:conclusion}
To meet the significant requirements of Quantity-of-Interest (QoI) preserving in error-bounded lossy compression and address the critical limitations of existing solutions, we propose QPET, a versatile and portable framework for error-bounded lossy compression that enables the preservation of a broad range of QoI while significantly improving the performance and quality of existing state-of-the-arts.
We integrate QPET into 3 representative error-bounded lossy compressors of different archetypes.
Experimental results demonstrate that QPET is faster than all existing QoI-preserving error-bounded lossy compression solutions, leading to 2x to 10x compression speedups over the baselines of both parameter-search methods and direct QoI-preserving methods. 
QPET also delivers up to 1000\% compression ratio improvements over existing QoI-preserving error-bounded lossy compression solutions.
In future research, we will optimize QPET in the following three aspects: (1) Enable the support for more QoI formats in practical use cases; 
(2) Optimize the strategies for determining the point-wise and global error bounds in QoI-preserving; 
(3) Enhance the throughput to further reduce the computational cost for QoI preservation;

\section{Acknowledgments}
This research was supported by the U.S. Department of Energy, Office of Science, Advanced Scientific Computing Research (ASCR), under contracts DE-AC02-06CH11357. This work was also supported by the National Science Foundation (Grant Nos. 2104023, 2311875, 2344717, OAC-2330367, OAC-2313122, OAC-2311756). We also acknowledge the computing resources provided by Advanced Cyberinfrastructure Coordination Ecosystem: Services \& Support (ACCESS) -- Purdue Anvil.

\onecolumn
\begin{multicols}{2}
\bibliographystyle{ACM-Reference-Format}

\bibliography{references}
\end{multicols}

\end{document}